\documentclass[aps,prd,a4paper,showpacs,nofootinbib,twocolumn]{revtex4-1}
\usepackage{hyperref}
\usepackage{ifpdf}
\usepackage{subfigure}
\usepackage{amsmath}
\usepackage{amssymb}
\usepackage{amsfonts}
\usepackage{epsf}
\usepackage{rotating}
\usepackage{graphicx}
\usepackage{amsmath}
\usepackage{fancyhdr}
\usepackage{lineno}
\usepackage{braket}
\usepackage{nicefrac}
\usepackage{graphics}
\usepackage{pstricks}
\usepackage{color}
\usepackage{multirow}
\usepackage{url}
\usepackage{empheq}

\def\sss{\scriptscriptstyle}

\allowdisplaybreaks
\begin{document}
	
\title{Non-leptonic beauty baryon decays and $CP$-asymmetries based on 
$SU(3)$-Flavor analysis}
\author{Shibasis Roy} 
\email{shibasisr@imsc.res.in}
\affiliation{The Institute of Mathematical Sciences, Taramani,
	Chennai 600113, India}
\affiliation{Homi Bhabha National Institute, Training School 
	Complex, Anushaktinagar, Mumbai  400094, India} 

\author{N.~G.~Deshpande}
\email{desh@uoregon.edu}
\affiliation{Institute of Theoretical Science, University of Oregon,
		Eugene, Oregon 94703, USA}

\author{Rahul Sinha} 
\email{sinha@imsc.res.in}
\affiliation{The Institute of Mathematical Sciences, Taramani,
		Chennai 600113, India}
\affiliation{Homi Bhabha National Institute, Training School Complex, 
		Anushaktinagar, Mumbai  400094, India}

	\date{\today}
	
\begin{abstract}
We consider hadronic weak decays of beauty-baryons into charmless baryons and 
pseudoscalar mesons in a general framework based on $SU(3)$ decomposition of 
the decay amplitudes. The advantage of the approach lies in the ability to perform an 
$SU(3)$ analysis of these decays without any particular set of dynamical 
assumptions while accounting for the effects of an arbitrarily broken $SU(3)$ 
flavor symmetry. Dictated by the symmetries of the effective Hamiltonian that allow us to relate or neglect reduced 
$SU(3)$ amplitudes, we derive several sum 
rule relations between amplitudes and relations between $CP$ asymmetries in 
these decays and identify those that hold even if $SU(3)$ is broken.
\end{abstract}
	
	%\pacs{12.60.Cn,14.70.Pw}
	
	\maketitle
%\abstract{ to be}
\section{Introduction}
LHCb is poised to collect a large data set of two-body weak decays of 
beauty-baryons~\cite{Bediaga:2012py,Bediaga:2018lhg,Aaij:2019pqz} into 
charmless baryons and pseudoscalar mesons paving the way to
a better understanding of heavy baryon decays. Significant progress has been 
made 
in the theoretical understanding of beauty meson 
decays~\cite{Zeppenfeld:1980ex,Savage:1989ub,Gronau:1990ka,Chau:1990ay,
Gronau:1994rj,Deshpande:1994pw,Gronau:1995hn,Gronau:1995hm,Grinstein:1996us,
Deshpande:1997rr,Deshpande:1997ar,Gronau:1998fn,Buras:1998ra,Beneke:2000ry,
He:2000dg,Deshpande:2000jp,Gronau:2000pk,Paz:2002ev,Dariescu:2002hw,
Wu:2002nz,Beneke:2003zv,Grossman:2003qp,Chiang:2003rb,Chiang:2004nm,
Buras:2004ub,Gronau:2006eb,Ali:2007ff,Cheng:2011qh,He:2013vta,Cheng:2014rfa,
Grinstein:2014aza,Hsiao:2015iiu,He:2018php,He:2018joe} spurred by the
experimental advances at flavor factories Belle and 
Babar~\cite{Bevan:2014iga,Kou:2018nap} as well as in 
LHCb~\cite{Bediaga:2012py,Bediaga:2018lhg,Aaij:2013fja,Aaij:2017ngy,
Aaij:2018tfw}.  The general framework of $SU(3)$ analysis in beauty mesons 
as well as charm meson decays~\cite{Hinchliffe:1995hz, 
Bhattacharya:2008ss,Pirtskhalava:2011va,Hiller:2012xm,Feldmann:2012js,Bhattacharya:2012ah, 
Grossman:2012ry,Gronau:2015rda,Muller:2015lua,Muller:2015rna,Cheng:2016ejf} 
into two pseudoscalars ($PP$), pseudoscalar-vector 
boson ($PV$), and two vector mesons ($VV$) has yielded several amplitude sum-rules and relationships between $CP$ asymmetries for various decay modes. While 
attempts have been made to analyze such decays for 
beauty-baryons, a comprehensive analysis is so far missing in the literature. 
In this paper, we consider the hadronic beauty-baryon decays into an octet or 
singlet of light baryons and a pseudoscalar meson based on the $SU(3)$ 
decomposition of the decay amplitudes approach pioneered for $B$-meson 
decays by Grinstein and Lebed~\cite{Grinstein:1996us}. In contrast to the 
methodology employed 
in~\cite{Lu:2009cm,Gronau:2013mza,Hsiao:2014mua,He:2015fwa,He:2015fsa,Zhu:2016bra,Hsiao:2017tif,Savage:1989qr,Pakvasa:1990if,Savage:1991wu,Lu:2016ogy,Geng:2017mxn,Geng:2018upx,Geng:2018bow,Geng:2018plk,Geng:2018rse,Grossman:2018ptn,Hsiao:2019yur,Geng:2019xbo,Wang:2019alu,Wang:2019dls,Jia:2019zxi} for
 bottom and charmed hadron decays,
our approach~\cite{Grinstein:1996us} facilitates an $SU(3)$ 
decomposition of the decays in terms of $SU(3)$-reduced amplitudes without any 
particular set of assumptions about the underlying dynamics.

The number of independent $SU(3)$-reduced amplitudes for any given initial and 
final state is exactly calculable and relations between decay amplitudes emerge 
naturally once the set of independent $SU(3)$-reduced amplitudes is smaller 
than the total number of possible decays. The counting of independent $SU(3)$ 
reduced amplitudes draws on 
the choice of the effective Hamiltonian, which in the most general case, 
indicate 44 independent reduced $SU(3)$ amplitudes equaling the number of all possible $\Delta S=-1$ and $\Delta S=0$ processes. In practice, the dimension-6 
effective Hamiltonian that mediates such hadronic decays of bottom baryons 
predict only 10 independent reduced $SU(3)$ amplitudes. One can therefore 
obtain amplitude relations between the decay modes that can be 
derived explicitly. Moreover, a systematic study of the $SU(3)$-breaking effects 
at the level of decay amplitudes, order by order expanded in the $SU(3)$ breaking 
parameter, is required to identify those amplitude relations that survive the $SU(3)$ breaking effects. Starting with the symmetries of the effective Hamiltonian, we relate or neglect reduced $SU(3)$ amplitudes to derive several sum rules relations between amplitudes and relations between $CP$ asymmetries while indicating more general relations that continue to hold when the $SU(3)$ symmetry is no longer exact. This study is crucial for a detailed analysis of the $CP$ asymmetry measurements in bottom baryons decays at the CDF and LHCb in recent times~\cite{Aaltonen:2011qt,Aaij:2012as,Aaltonen:2014vra,Aaij:2016cla,Aaij:2017pgy,Aaij:2018lsx,Aaij:2018tlk,Aaij:2019rkf}. 
  
The approach to decompose the decay amplitudes in terms of reduced $SU(3)$ 
amplitudes is presented in Sec.~\ref{sec:t}. 
The relation between the $SU(3)$ Clebsch-Gordon (CG) coefficients in terms of the isoscalar factors and the $SU(2)$ CG coefficients is outlined in Appendix~\ref{sec:App0}. The results are summarized in Appendix.~\ref{sec:App1} and \ref{sec:App2}. In Sec.~\ref{sec:t1} we perform the $SU(3)$ decomposition of unbroken effective hadronic weak decay Hamiltonian. The relations between the amplitudes for beauty baryon decays into octets of light baryons and pseudoscalar mesons are derived in Sec.~\ref{sec:relations}. The effects of $SU(3)$ breaking on account of $s$-quark mass are considered in Sec.~\ref{sec:SU3}. The corresponding relations between $CP$ asymmetries are derived in Sec.~\ref{sec:CP}. We finally conclude in Sec.~\ref{sec:conclusion}.

%In Appendix~\ref{sec:App3} we have included the relevant meson and baryon wave 
%functions.

\section{Application of $SU(3)$ to decay amplitudes}
\label{sec:t}

The $SU(3)$ decomposition of physical amplitudes describing a decay process 
involves writing it in terms of reduced matrix elements of explicit $SU(3)$ 
operators with appropriate coefficients. The procedure is a 
straightforward application of Wigner-Eckart theorem for the group $SU(3)$ 
where the reduced matrix elements are all possible $SU(3)$ invariants with 
Clebsch-Gordon (CG) coefficients connecting the basis involving physical 
states to the group theoretic basis.  

The most general Hamiltonian ${\cal H}$ which connects~\cite{Grinstein:1996us} the initial and 
final states via the matrix elements $\braket{f|{\cal H}|i}$, consists of 
exactly those representations $\mathbf{R}$ 
appearing in $\mathbf{f}\otimes\mathbf{\bar{i}}$, where the labels $i$ and $f$ 
denote both physical states and $SU(3)$ representations. It is important to note that in addition to the usual $SU(3)$ CG coefficients that arise from coupling 
$\mathbf{f}\otimes\mathbf{\overline{i}}$, the most general effective 
Hamiltonian ($\mathcal{H}$) itself involves unknown coefficients 
appearing in front of every $SU(3)$ representation. A priori these 
coefficients are all independent of each other which get determined once 
a particular form of effective Hamiltonian is assumed. The 
states of $SU(3)$ representations are uniquely distinguished when in addition to the $I_3$ and $Y$ values, the isospin 
Casimir $I^2$ is also specified. The full reduced $SU(3)$ amplitude is thus 
described by $\braket{f||{R_I}||i}$.
The expression of the amplitudes in terms of reduced $SU(3)$ amplitudes 
is concisely given as,
\begin{widetext}
\vspace{-0.2cm}
	\begin{eqnarray}
	\label{Master Formula}
	\mathcal{A}(i \to f_{b} 
	f_{m})=(-1)^{I_{3}+\frac{Y}{2}+\frac{T}{3}}\quad\sum_{\mathclap 
	{\substack{\{f,	\, R\} \\
				Y^{b}+Y^{m}=Y^{f}, \, Y^{f}-Y^{i}=Y^{H}\\
				I_{3}^{b}+I_{3}^{m}=I_{3}^{f},\, 
				I_{3}^{f}-I_{3}^{i}=I_{3}^{H}}}} 
	C^{I_{3}^{b}\, I_{3}^{m}\, I_{3}^{f} }_{I^{b}\, I^{m}\, I^{f}} 
	\quad \begin{pmatrix}
	\mathbf{f_{b}}       & \mathbf{f_{m}}  & \mathbf{f} \\
	\noalign{\smallskip}    
	(Y^{b},I^{b},I^{b}_{3}) & (Y^{m},I^{m},I^{m}_{3}) 	& 
	(Y^{f},I^{f},I^{f}_{3}) \\
	\end{pmatrix} \nonumber\\
	\quad \begin{pmatrix}
	\mathbf{f}       & \mathbf{\bar{i}}  & \mathbf{R} \\
	\noalign{\smallskip}    
	(Y^{f},I^{f},I^{f}_{3}) & (-Y^{i},I^{i},-I^{i}_{3}) 	& 
	(Y^{H},I^{H},I^{H}_{3}) \\
	\end{pmatrix}  C^{I_{3}^{f}\, -I_{3}^{i}\, I_{3}^{H} }_{I^{f}\, I^{i}\, 
	I^{H}} 
	\langle \mathbf{f}\parallel \mathbf{R_{I}} \parallel 
	\mathbf{i}\rangle,
	\end{eqnarray}     
\end{widetext}
where, $C^{a,b,c}_{A,B,C}$ are the $SU(2)$ Clebsch-Gordon coefficients and 
\begin{equation}
\begin{pmatrix}
\mathbf{R_{a}}       & \mathbf{R_{b}}  & \mathbf{R_{c}} \\
\noalign{\smallskip}    
(Y^{a},I^{a},I^{a}_{3}) & (Y^{b},I^{b},I^{b}_{3}) 	& 
(Y^{c},I^{c},I^{c}_{3}) 
\end{pmatrix} .
\end{equation}
are the $SU(3)$ isoscalar coefficients obtained by coupling the 
representations $\mathbf{R_{a}}\otimes\mathbf{R_{b}} \to \mathbf{R_{c}}$. $T$ 
is the triality of a $SU(3)$ representation\footnote{\scriptsize{For an 
$SU(3)$-representation with $m$ and $n$ fundamental and anti-fundamental 
indices, the triality of the representation is given by 
$T\big((m,n)\big)=(m-n)\, \textbf{mod} \, 3$}} that ensures the reality of the 
phase appearing in Eq.~\eqref{Master Formula}. The symmetry properties of the 
$SU(3)$ isoscalar factor and its role in obtaining the $SU(3)$ CG 
coefficients~\cite{Shortley,deSwart:1963pdg,Samios:1974tw,Kaeding:1995re,Kaeding:1995vq,Grinstein:2004kn}
 is outlined in Appendix~\ref{sec:App0}. The amplitude is written with 
specific attention to the order in which the representations are coupled, the 
final state representations are coupled via $\textbf{f}_{b}\otimes \textbf{f}_{m} 
\to \textbf{f} $, where the product ($f$) is then coupled through the conjugate of the initial representation or equivalently $\mathbf{f}\otimes\bar{\mathbf{i}}\to {\cal H}$. This ensures that all possible $SU(3)$ representations are indeed generated in case of the most general effective Hamiltonian.
Given a form of effective Hamiltonian ($\mathcal{H}_{\text{eff}}$), it 
can be $SU(3)$ decomposed, 
\begin{align}
\label{eff Ham definite}
\mathcal{H}_{\text{eff}}=\sum_{\substack{ \{Y,I,I_{3}\}\\ 
\mathbf{R}}}\mathcal{F}^{\{Y,I,I_{3}\}}_{\mathbf{R}} \mathbf{R_{I}},
\end{align}
where $\mathcal{F}^{\{Y,I,I_{3}\}}_{\mathbf{R}}$ depends on the $SU(3)$ CG 
coefficients appearing in front of the $SU(3)$ representations 
($\mathbf{R_{I}}$). Moreover $\mathcal{F}^{\{Y,I,I_{3}\}}_{\mathbf{R}}$ also 
contains additional factors entering Eq.~\eqref{eff Ham definite} in form of 
Wilson coefficients and CKM elements. It is also important to note that by 
knowing the dynamical coefficients for different isospin values in a given 
$SU(3)$ representation, one can drop the isospin Casimir label ($I$) and 
express the Wigner-Eckart reduced matrix element $\langle \mathbf{f}\parallel 
\mathbf{R} \parallel \mathbf{i}\rangle$, in its usual form, independent of the 
isospin $I$ 
label. By using completeness of $SU(3)$ CG coefficients up to a phase factor, 
\begin{align}
\langle \mathbf{f}\parallel \mathbf{R_{I}} \parallel 
\mathbf{i}\rangle=\underbrace{\mathcal{F}^{\{Y,I,I_{3}\}}_{\mathbf{R}} 
\sqrt{\frac{\text{dim f }}{\text{dim R }}}}_{\text{dynamical Coeff.
of}~\mathcal{H}} \langle \mathbf{f}\parallel \mathbf{R} \parallel 
\mathbf{i}\rangle.
\end{align}
Alternatively, one can directly start with the given form of effective 
Hamiltonian in Eq.~\eqref{eff Ham definite} and perform an $SU(3)$ 
decomposition of the decay amplitude;
\begin{widetext}
\begin{eqnarray}
\label{eq:Amp2}
	\mathcal{A}(i \to f_{b} f_{m})=\sum_{\substack{ \{Y^{H},I^{H},I^{H}_{3}\}\\ 
	\mathbf{R}}}\mathcal{F}^{\{Y,I,I_{3}\}}_{\mathbf{R}} 
	\sum_{\mathclap{\substack{\{f\} \\
				Y^{b}+Y^{m}=Y^{f}, \, Y^{f}=Y^{i}+Y^{H}\\
				I_{3}^{b}+I_{3}^{m}=I_{3}^{f},\, 
				I_{3}^{f}=I_{3}^{i}+I_{3}^{H}}}} C^{I_{3}^{b}\, I_{3}^{m}\, 
				I_{3}^{f} }_{I^{b}\, I^{m}\, I^{f}} \quad \begin{pmatrix}
	\mathbf{f_{b}}       & \mathbf{f_{m}}  & \mathbf{f} \\
	\noalign{\smallskip}    
	(Y^{b},I^{b},I^{b}_{3}) & (Y^{m},I^{m},I^{m}_{3}) 	& 
	(Y^{f},I^{f},I^{f}_{3}) \\
	\end{pmatrix} \nonumber\\
	\quad \begin{pmatrix}
	\mathbf{R}       & \mathbf{i}  & \mathbf{f} \\
	\noalign{\smallskip}    
	(Y^{H},I^{H},I^{H}_{3}) & (Y^{i},I^{i},I^{i}_{3}) 	& 
	(Y^{f},I^{f},I^{f}_{3}) \\
	\end{pmatrix}  C^{I_{3}^{H}\, I_{3}^{i}\, I_{3}^{f} }_{I^{H}\, I^{i}\, 
		I^{f}} \langle \mathbf{f}\parallel \mathbf{R} \parallel 
	\mathbf{i}\rangle.
\end{eqnarray}   
\end{widetext}

%The order in which 
%the representation of the final states $\textbf{f}_{b}\otimes 
%\textbf{f}_{m} 
%\to \textbf{f} $ is coupled to the interaction Hamiltonian is 
%$\textbf{R}$ is 
%$\textbf{f}\otimes \overline{\textbf{i}} \to \textbf{R} $.
%
%Another way of working would be to directly start with a given form of 
%effective Hamiltonian that can be decomposed under $SU(3)$; 
%
%

The case of our interest, namely, $B_b(\overline{\textbf{3}})\to 
B(\textbf{8})\, 
P(\textbf{8})$, where $B_b$, the initial anti-triplet ($\overline{\textbf{3}}$) 
beauty-baryon undergoes a charmless decay into an octet baryon ($B$) and an 
octet 
pseudoscalar meson ($P$), is described by a Hamiltonian with $\Delta Q=0$ and 
$\Delta S$ (equivalent to $\Delta I_{3}$ and $\Delta Y$ representation). The 
possible decays can be divided into two sub classes, namely the $\Delta S=0$ 
and $\Delta S=-1$ transitions. The allowed final state $SU(3)$ representations 
($\mathbf{f}$) are; $\mathbf{1}$, $\mathbf{8}_{1}$, $\mathbf{8}_{2}$, 
$\mathbf{10}$, $\overline{\mathbf{10}}$, $\mathbf{27}$. There are 22 physical 
process possible for $\Delta S=0$ and another 22 for $\Delta S=-1$. In 
Appendix~\ref{sec:App1} and Appendix~\ref{sec:App2} respectively each of these 
decay modes are decomposed in terms of the $SU(3)$ reduced amplitudes that add upto 
44. Since the physical $\eta$ and $\eta^{'}$ mesons are admixtures of octet 
$\eta_{8}$ and singlet $\eta_{1}$ mesons, a study of 
$B_{b}(\overline{\textbf{3}})\to B(\textbf{8})\, 
P(\textbf{1})$ is also necessary. Therefore one has to take into account 8 (4 
each 
for $\Delta S=-1$ and $\Delta S=0$) additional independent $SU(3)$ amplitudes 
which are also
described in Appendix~\ref{sec:App1} and Appendix~\ref{sec:App2}.  

We emphasize that this way of counting accounts for a complete set of reduced 
amplitudes, regardless of the specific form of interaction Hamiltonian. In 
particular, this decomposition holds even if the $SU(3)$ symmetry is arbitrarily 
broken and there is no physical reason to organize particles in $SU(3)$ 
multiplets. At this point, every process is independent and to find relations 
among them requires assuming a specific form of the interaction Hamiltonian.

%%%%%%%%%%%%%%%%%%%%%%%%%%%%%%%%%%%%%%%%%%
\section{$SU(3)$ decomposition of unbroken effective Hamiltonian}
\label{sec:t1}
The lowest order effective Hamiltonian~\cite{Ciuchini:1993vr,Buchalla:1995vs,Buras:1998raa}
for charmless $b$-baryon decays consists $\Delta S=-1$ and $ \Delta S=0$ parts. Each part is composed from the 
operators $ Q_{1}$,\ldots, $ Q_{10}$. The complete Hamiltonian can be written 
as:
\begin{widetext}
\begin{eqnarray}
\label{eff H}
\mathcal{H}_{\text{eff}}=\frac{4G_{F}}{\sqrt{2}}\Big[\lambda^{(s)}_{u}
\Big(C_{1}(Q^{(u)}_{1}-Q^{(c)}_{1})+C_{2}(Q^{(u)}_{2}-Q^{(c)}_{2})\Big)-
\lambda^{(s)}_{t} \sum_{i=1,2}C_{i}Q^{(c)}_{i}-
\lambda^{(s)}_{t}
\sum_{i=3}^{10}C_{i}Q_{i}^{(s)}\nonumber \\ +\lambda^{(d)}_{u}
\Big(C_{1}(Q^{(u)}_{1}-Q^{(c)}_{1})+C_{2}(Q^{(u)}_{2}-Q^{(c)}_{2})\Big)-\lambda^{(d)}_{t} \sum_{i=1,2}C_{i}Q^{(c)}_{i}-
\lambda^{(d)}_{t}
\sum_{i=3}^{10}C_{i}Q_{i}^{(d)} \Big],
\end{eqnarray} 
\end{widetext}
where $V_{ub}V_{us}^{*}=\lambda_{u}^{s}$, $V_{ub}V_{ud}^{*}=\lambda_{u}^{d}$, 
$V_{tb}V_{ts}^{*}=\lambda_{t}^{s}$, $V_{tb}V_{td}^{*}=\lambda_{t}^{d}$ are the 
CKM elements and $C_{i}$ s are the Wilson coefficients. $ Q_{1}$ and $ Q_{2}$ 
are the ``Tree'' operators:
\begin{align}
Q_{1}^{(u)} & =  (\overline{u}^{i}_{L}\gamma^{\mu}b^{j}_{L})
(\overline{s}^{j}_{L}\gamma_{\mu}u^{i}_{L}) \nonumber \\
Q_{1}^{(c)} & =  (\overline{c}^{i}_{L}\gamma^{\mu}b^{j}_{L})
(\overline{s}^{j}_{L}\gamma_{\mu}c^{i}_{L}) \nonumber\\
Q_{2}^{(u)}&=(\overline{u}^{i}_{L}\gamma^{\mu}b^{i}_{L})
(\overline{s}^{j}_{L}\gamma_{\mu}u^{j}_{L})  \nonumber  \\
Q_{2}^{(c)}&=(\overline{c}^{i}_{L}\gamma^{\mu}b^{i}_{L})
(\overline{s}^{j}_{L}\gamma_{\mu}c^{j}_{L}). 
\end{align}
$ Q_{3}$, \ldots, $ Q_{6}$ are the ``Gluonic Penguin'' operators:
\begin{align}
Q_{3}^{(s)} & = (\overline{s}^{i}_{L}\gamma^{\mu}b^{i}_{L})
\sum_{q=u,d,s}(\overline{q}^{j}_{L}\gamma_{\mu}q^{j}_{L})\nonumber \\
Q_{4}^{(s)} & = (\overline{s}^{i}_{L}\gamma^{\mu}b^{j}_{L}) 
\sum_{q=u,d,s}(\overline{q}^{j}_{L}\gamma_{\mu}q^{i}_{L})\nonumber \\
Q_{5}^{(s)} & = (\overline{s}^{i}_{L}\gamma^{\mu}b^{i}_{L})
\sum_{q=u,d,s}(\overline{q}^{j}_{R}\gamma_{\mu}q^{j}_{R})\nonumber \\
Q_{6}^{(s)} & = (\overline{s}^{i}_{L}\gamma^{\mu}b^{j}_{L})
\sum_{q=u,d,s}(\overline{q}^{j}_{R}\gamma_{\mu}q^{i}_{R}). 
\end{align}
Out of the four ``EWP'' (i.e. ``Electroweak Penguins'') Operators: 
$ Q_{7}$, \ldots, $ Q_{10}$,  $ Q_{7}$ and $ Q_{8}$:
\begin{align}
Q_{7}^{(s)} & = \frac{3}{2} (\overline{s}^{i}_{L}\gamma^{\mu}b^{i}_{L})
\sum_{q=u,d,s}e_{q}(\overline{q}^{j}_{R}\gamma_{\mu}q^{j}_{R}),\nonumber \\ 
Q_{8}^{(s)} & =\frac{3}{2} (\overline{s}^{i}_{L}\gamma^{\mu}b^{j}_{L})
\sum_{q=u,d,s}e_{q}(\overline{q}^{j}_{R}\gamma_{\mu}q^{i}_{R}),
\end{align}
are typically ignored in hadronic decays because of the smallness of $ C_{7}$ 
and  $C_{8}$ with respect to the other Wilson Coefficients.\\
The remaining ``EWP'' operators are:
\begin{align}
Q_{9}^{(s)} & = \frac{3}{2} (\overline{s}^{i}_{L}\gamma^{\mu}b^{i}_{L})
\sum_{q=u,d,s}e_{q}(\overline{q}^{j}_{L}\gamma_{\mu}q^{j}_{L}),\nonumber \\
Q_{10}^{(s)} & =\frac{3}{2} (\overline{s}^{i}_{L}\gamma^{\mu}b^{j}_{L})
\sum_{q=u,d,s}e_{q}(\overline{q}^{j}_{L}\gamma_{\mu}q^{i}_{L}).
\end{align}

\noindent ${\cal H}_{\rm eff}$  is a  linear combinations of four quark operators of the form $(\overline{q}_{1}b)(\overline{q_{2}}q_{3})$.
These operators transform as $3\bigotimes 3 \bigotimes \overline{3}$
under $SU(3)$-flavor and can be decomposed into sums of irreducible  operators corresponding to irreducible $SU(3)$ representations:
$\textbf{15}\,$,$\,\mathbf{\overline{6}}\,$,$\,\mathbf{3^{(6)}}\,$,$\,\mathbf{3^{(\bar{3})}}\,$ where  the superscript index: `6' (`$\overline{3}$') indicates the origin of $\mathbf{3}$ out of the two possible representations arising from the tensor product of $q_{1}$ and $q_{2}$. The $SU(3)$ triplet representation of quarks ($q_{i}$) and its conjugate denoting the anti-quarks ($\overline{q_{i}}$) consist of the flavor states;
\begin{equation}
\label{eqn:q-antiq}
q_{i}=
\begin{pmatrix}
& u &\\& d &\\& s &\\ 
\end{pmatrix}
\qquad
\overline{q_{i}}=
\begin{pmatrix}
&\overline{d}& \\ &-\overline{u}& \\ &\overline{s}& \\ 
\end{pmatrix}
\end{equation}
According to the sign convention chosen in Eq.~\eqref{eqn:q-antiq}, the meson  
wavefunctions are given as, 
\begin{eqnarray}
K^{+}=u\overline{s}, \qquad K^{-}=-s\overline{u}, \qquad K^{0}= d\overline{s}, 
\qquad  \overline{K}^{0}=s\overline{d}\qquad\nonumber\\
\pi^{+}=u\overline{d},\qquad \pi^{-}=-d\overline{u},\qquad 
\pi^{0}=\frac{1}{\sqrt{2}}(d\overline{d}-u\overline{u})\qquad\nonumber\\
\eta_{8}=-\frac{1}{2\sqrt{6}}(u\overline{u}+d\overline{d}-2s\overline{s})\quad
\eta_{1}=-\frac{1}{\sqrt{3}}(u\overline{u}+d\overline{d}+s\overline{s})\quad\nonumber
\end{eqnarray}
The physical mesons $\eta,\, \eta^{'}$ are related to the $\eta_{8}$ and 
$\eta_{1}$ through the $SO(2)$ rotation,
\begin{equation}
\begin{pmatrix}
\eta\\
\eta^{\prime}
\end{pmatrix}=\begin{pmatrix}
-\cos\theta & \sin \theta\\
-\sin\theta & -\cos \theta
\end{pmatrix}
\begin{pmatrix}
\eta_{8}\\
\eta_{1}
\end{pmatrix}
\end{equation}
where the definition of $\theta$ is consistent with the overall notation for the meson wavfunctions as well as agreeing with the phenomenologically determined value of $\theta$.  
In the following table the four quark operators, which appear in ${\cal H}_{\rm 
eff}$,  are decomposed using $SU(3)$ Clebsch-Gordan tables.  It is worthwhile to note that in the Hamiltonian operators appear as  $\overline{q}_{1}\,\overline{q_{2}}\, 
q_{3}$ whereas in Table~\ref{Tab 1} they are expressed conveniently as $q_{1}\,q_{2}\overline{q}_{3}$. With the help of Table \ref{Tab 1}, the effective Hamiltonian can be expressed 
in terms of operators having definite $SU(3)$ transformation properties. In 
particular, the tree, gluonic and electroweak penguin part of the 
effective Hamiltonian consist of~\cite{Paz:2002ev}, 
\begin{widetext}

\begin{table*}[ht!]
\begin{tabular}{|c|c|c|c|c|c|c|c|c|c|c|}	
\hline
\,&$15_{I=1}$&$15_{I=0}$&$\overline{6}_{I=1}$&$3^{(6)}_{I=0}$&$3^{(\overline{3})}_{I=0}$&$15_{I=3/2}$&$15_{I=1/2}$&$\overline{6}_{I=1/2}$&$3^{(6)}_{I=1/2}$&$3^{(\overline{3})}_{I=1/2}$
 \\ 
\hline
$u\,s\,\overline{u}$&$-1/2$&$-1/\sqrt{8}$&$-1/2$&$-1/\sqrt{8}$&$-1/2$&&&&& \\ \hline
$s\,u\,\overline{u}$&$-1/2$&$-1/\sqrt{8}$&$1/2$&$-1/\sqrt{8}$&$1/2$&&&&& \\ \hline
$s\,d\,\overline{d}$&$1/2$&$-1/\sqrt{8}$&$-1/2$&$-1/\sqrt{8}$&$1/2$&&&&& \\ \hline
$d\,s\,\overline{d}$&$1/2$&$-1/\sqrt{8}$&$1/2$&$-1/\sqrt{8}$&$-1/2$&&&&& \\ \hline
$s\,s\,\overline{s}$&&$1/\sqrt{2}$&&$-1/\sqrt{2}$&&&&&& \\ \hline
$u\,d\,\overline{u}$&&&&&&$-1/\sqrt{3}$&$-1/\sqrt{24}$&$1/2$&$-1/\sqrt{8}$&$-1/2$ \\ \hline
$d\,u\,\overline{u}$&&&&&&$-1/\sqrt{3}$&$-1/\sqrt{24}$&$-1/2$&$-1/\sqrt{8}$&$1/2$ \\ \hline
$d\,d\,\overline{d}$&&&&&&$1/\sqrt{3}$&$-1/\sqrt{6}$&&$-1/\sqrt{2}$& \\ \hline
$d\,s\,\overline{s}$&&&&&&&$\sqrt{3}/\sqrt{8}$&$1/2$&$-1/\sqrt{8}$&$1/2$ \\ \hline
$s\,d\,\overline{s}$&&&&&&&$\sqrt{3}/\sqrt{8}$&$-1/2$&$-1/\sqrt{8}$&$-1/2$ \\ \hline
\end{tabular}
\caption{Operator Decomposition}
\label{Tab 1}
\end{table*}

\begin{align}
\label{T Ham}
\frac{\sqrt{2}\mathcal{H}_{\sss{\rm T}}}{4G_{F}}=&\left\{\lambda_{u}^{s}
\left[\frac{(C_{1}+C_{2})}{2}\left(-\textbf{15}_{1}-\frac{1}{\sqrt{2}}
\textbf{15}_{0}-\frac{1}{\sqrt{2}}\textbf{3}^{(\textbf{6})}_{0}\right)
+\frac{(C_{1}-C_{2})}{2}\left(\textbf{6}_{1}+\textbf{3}^{(\overline
	{\textbf{3}})}_{0}\right)\right]\right.\nonumber
\\&+\left.\lambda_{u}^{d}\left[\frac{(C_{1}+C_{2})}{2}\left(-\frac{2}{\sqrt{3}}
\textbf{15}_{3/2}-\frac{1}{\sqrt{6}}\textbf{15}_{1/2}-\frac{1}{\sqrt{2}}
\textbf{3}^{(\textbf{6})}_{1/2}\right)+\frac{(C_{1}-C_{2})}{2}
\left(-\textbf{6}_{1/2}+\textbf{3}^{(\overline{\textbf{3}})}_{1/2}\right)\right]
\right\},
\end{align} 

\begin{align}
\label{P g Ham}
\frac{\sqrt{2}\mathcal{H}_{\sss{\rm g}}}{4G_{F}}=&\left\{-\lambda_{t}^{s}
\left[-\sqrt{2}(C_{3}+C_{4})\textbf{3}^{(\textbf{6})}_{0}+(C_{3}-C_{4})
\textbf{3}^{(\overline{\textbf{3}})}_{0}\right]\right. 
-\left.\lambda_{t}^{d}\left[-\sqrt{2}(C_{3}+C_{4})\textbf{3}^{(\textbf{6})}_{1/2}
+(C_{3}-C_{4})\textbf{3}^{(\overline{\textbf{3}})}_{1/2}\right]\right.\nonumber
\\&-\left.\lambda_{t}^{s}
\left[-\sqrt{2}(C_{5}+C_{6})\textbf{3}^{(\textbf{6})}_{0}+(C_{5}-C_{6})
\textbf{3}^{(\overline{\textbf{3}})}_{0}\right]\right. 
-\left.\lambda_{t}^{d}\left[-\sqrt{2}(C_{5}+C_{6})\textbf{3}^{(\textbf{6})}_{1/2}
+(C_{5}-C_{6})\textbf{3}^{(\overline{\textbf{3}})}_{1/2}\right]\right\},
\end{align} 

\begin{align}
\label{P EW Ham}
\frac{\sqrt{2}\mathcal{H}_{\sss{\rm EWP}}}{4G_{F}}=&\left\{-\lambda_{t}^{s}
\left[\frac{(C_{9}+C_{10})}{2}\left(-\frac{3}{2}\textbf{15}_{1}
-\frac{3}{2\sqrt{2}}\textbf{15}_{0}+\frac{1}{2\sqrt{2}}
\textbf{3}^{(\textbf{6})}_{0}\right)+\frac{(C_{9}-C_{10})}{2}
\left(\frac{3}{2}\textbf{6}_{1}+\frac{1}{2}
\textbf{3}^{(\overline{\textbf{3}})}_{0}\right)\right]\right.\nonumber
 \\ &\left.-\lambda_{t}^{d}\left[\frac{(C_{9}+C_{10})}{2}\left(-\sqrt{3}\, 
\textbf{15}_{3/2}-\frac{1}{2}\sqrt{\frac{3}{2}}\textbf{15}_{1/2}
+\frac{1}{2\sqrt{2}}\textbf{3}^{(\textbf{6})}_{1/2}\right)
+\frac{(C_{9}-C_{10})}{2}\left(-\frac{3}{2}\textbf{6}_{1/2}
+\frac{1}{2}\textbf{3}^{(\overline{\textbf{3}})}_{1/2}\right)\right]\right\}.
\end{align} 
\end{widetext}
It is clear from Table~\ref{Tab 1} that higher $SU(3)$ representations like 
$\mathbf{24}$, $\mathbf{42}$ and $\mathbf{15^{'}}$  are absent in the unbroken 
Hamiltonian.
%%%%%%%%%%%%%%%%%%%%%%%%%%%%%%%%%%%%%%%%%%%%%%%%%%%%%%%%%%%%%%
\section{Amplitude relations}
\label{sec:relations}
From the tree and Electroweak part of the Hamiltonian one can project out the 
coefficients corresponding to the $\mathbf{15}$ part of the Hamiltonian and 
write down the following relations between reduced matrix elements regardless 
of the initial and final states
\begin{align}
%\frac{\langle \mathbf{f}\parallel \mathbf{15_{0}} \parallel 
%\mathbf{i}\rangle}{ 
%\langle \mathbf{f}\parallel \mathbf{15_{1}} \parallel 
%\mathbf{i}\rangle}=\frac{1}{\sqrt{2}},\qquad 
\frac{\braket{\mathbf{f}\parallel \! \mathbf{15_{0}} \!\parallel 
\mathbf{i}}}{\braket{\mathbf{f}\parallel \!\mathbf{15_{1}}\!\parallel
	\mathbf{i}}}=\frac{1}{\sqrt{2}},\qquad 
\frac{\langle \mathbf{f}\parallel \!\mathbf{15_{\sss \frac{\mathbf{1}}{\mathbf 
2}}}\!\parallel 
\mathbf{i}\rangle}{ \langle \mathbf{f}\!\parallel 
\mathbf{15_{\sss \frac{\mathbf{3}}{\mathbf 2}}}\! \parallel 
\mathbf{i}\rangle}=\frac{1}{2\sqrt{2}}\nonumber,\\
\frac{\lambda^{d}_{t}}{\lambda^{s}_{t}}\frac{\langle \mathbf{f}\!\parallel 
\mathbf{15_{0}} \!\parallel \mathbf{i}\rangle_{{\sss{\rm EWP}}}}{ \langle 
\mathbf{f}\!\parallel \mathbf{15}_{\sss \frac{\mathbf{1}}{\mathbf 
2}}\!\parallel 
\mathbf{i}\rangle_{{\sss {\rm \mathrm{EWP}}}}}=\sqrt{3},~
\frac{\lambda^{d}_{u}}{\lambda^{s}_{u}}\frac{\langle \mathbf{f}\!\parallel 
\mathbf{15_{0}} \!\parallel \mathbf{i}\rangle_{{\sss{\rm T}}}}{ \langle 
\mathbf{f}\!\parallel \mathbf{15_{\sss \frac{\mathbf{1}}{\mathbf 2}}}\! 
\parallel 
\mathbf{i}\rangle_{{\sss{\rm T}}}}=\sqrt{3}
\end{align} 
In case of several different operator structures contributing to the 
Hamiltonian as is the case in Eq.~\eqref{eff H}, the relations between reduced 
matrix elements are expressed in the following way,
\begin{align}
\frac{\langle \mathbf{f}\parallel \mathbf{R_{I}} \parallel \mathbf{i}\rangle}{ \langle \mathbf{f}\parallel \mathbf{R_{I^{'}}} \parallel \mathbf{i}\rangle}=\frac{\sum_{l}\mathcal{C}_{l}C_{l}}{\sum_{m}\mathcal{C}_{m}C^{'}_{m}},
\end{align}  
where, the $C^{(')}_{i}$ are the coefficients of the different components of 
the Hamiltonian and  $\mathcal{C}_{j}$'s are the CG coefficients and the sums extend over 
all the corresponding contributions to the Hamiltonian. In addition, the 
absence of some of the $SU(3)$ representations in the Hamiltonian is a 
consequence of the vanishing dynamical coefficients corresponding to the  
reduced matrix elements $\braket{\mathbf{f}\parallel \mathbf{42} \parallel 
\mathbf{i}}$, 
$\braket{\mathbf{f}\parallel \mathbf{24} \parallel \mathbf{i}}$ and 
$\braket{\mathbf{f}\parallel \mathbf{15^{'}} \parallel \mathbf{i}}$,
regardless of the $I$ value and initial and final states.  The $\Delta 
S=-1$ and $\Delta S=0$ decay amplitudes and the reduced $SU(3)$ elements are 
expressed as column matrices $\mathcal{A}$ and $\mathcal{R}$ respectively and 
related by the matrix equation 
\begin{equation}
\mathcal{A}=T\mathcal{R},
\end{equation}
where $T$ is the coefficient matrix related to the tree part of the Hamiltonian 
described in Eq.~\eqref{T Ham}. The rank of matrix $T$ is lower than the total 
number 
of decay modes suggesting that not all of the reduced $SU(3)$ matrix elements are 
independent. The number of actually independent reduced $SU(3)$ matrix elements 
are equal to the rank of matrix $T$. The number of amplitude relations can now 
be estimated unambiguously which is the difference between the total number 
decay modes and rank of $T$. Similar exercise is performed for the penguins 
where the coefficient matrix is given by $P_{g}$ and $P_{EW}$. It is 
advantageous to factor out the CKM elements $\lambda_{u,t}^{s,d}$ and 
write the decay amplitude in terms of tree and penguin reduced amplitudes, 
\begin{eqnarray}
\label{SP wave}
\mathcal{A}^{\mathcal{S}}=&\lambda_{u}^{q}\mathcal{A}_{\text{T}}^{\mathcal{S}}+
\lambda_{t}^{q}\mathcal{A}_{\text{P}}^{\mathcal{S}},\nonumber\\
\mathcal{A}^{\mathcal{P}}=&\lambda_{u}^{q}\mathcal{A}_{\text{T}}^{\mathcal{P}}+
\lambda_{t}^{q}\mathcal{A}_{\text{P}}^{\mathcal{P}},
\end{eqnarray}
where $q=s,\,d$ denote the $\Delta S=-1,\,0$ process, $\mathcal{S}$ and 
$\mathcal{P}$ denote the $S$ wave and $P$ wave amplitudes of the decay. Now the 
coefficient matrix can be rewritten where the entries are nothing but products 
of Wilson Coefficients ($C_{i}$) and Clebsch-Gordon coefficients. Of course, the number of independent rows 
remain unchanged and the matrix equations take the form,
\begin{equation}
\mathcal{A}_{T}=\mathcal{T}\mathcal{R} \qquad \qquad \mathcal{A}_{P}=\mathcal{P}\mathcal{R}
\end{equation}
for both the $\mathcal{S}$-wave and the $\mathcal{P}$-wave part.
At this point, it is important to recall the penguin part of the Hamiltonian 
described in Eqs.~\eqref{P g Ham} and \eqref{P EW Ham}. In case of gluonic 
penguins, 
$\overline{\mathbf{6}}$ and \textbf{15} of $SU(3)$ are absent which result in a 
smaller set of independent reduced $SU(3)$ matrix elements. This implies 
additional amplitude relations between decay modes, some of which are broken 
once the Electroweak penguins are taken into account in the unbroken 
Hamiltonian. We include Electroweak penguins that have parts transforming as 
\textbf{3}, $\overline{\mathbf{6}}$ and \textbf{15} of $SU(3)$ and retain all the 
reduced $SU(3)$ matrix elements. As a result, the amplitude relations derived 
hold for the gluonic penguin part as well as the Electroweak penguin part of 
the unbroken Hamiltonian. We begin with identifying the identical rows of the 
$\mathcal{T}$ and  $\mathcal{P}$ matrices which readily gives the simplest 
amplitude relations for the tree part
\begin{align}
\label{T Rel1}
\mathcal{T}(\Lambda_{b}^{0}\to \Sigma^{-}K^{+})&=\mathcal{T}(\Xi_{b}^{0}\to 
\Xi^{-}\pi^{+}),\\
\mathcal{T}(\Lambda_{b}^{0}\to p^{+}\pi^{-})&=\mathcal{T}(\Xi_{b}^{0}\to 
\Sigma^{+}K^{-}),\\
\mathcal{T}(\Xi_{b}^{-}\to nK^{-})&=\mathcal{T}(\Xi_{b}^{-}\to 
\Xi^{0}\pi^{-}),\\
\mathcal{T}(\Xi_{b}^{-}\to \Xi^{-}K^{0})&=\mathcal{T}(\Xi_{b}^{-}\to 
\Sigma^{-}\overline{K}^{0}),\\
\mathcal{T}(\Xi_{b}^{0}\to \Xi^{-}K^{+})&=\mathcal{T}(\Lambda_{b}^{0}\to 
\Sigma^{-}\pi^{+}),\\
\mathcal{T}(\Xi_{b}^{0}\to \Sigma^{-}\pi^{+})&=\mathcal{T}(\Lambda_{b}^{0}\to 
\Xi^{-}K^{+}),\\
\mathcal{T}(\Xi_{b}^{0}\to \Sigma^{+}\pi^{-})&=\mathcal{T}(\Lambda_{b}^{0}\to 
p^{+}K^{-}),\\
\mathcal{T}(\Xi_{b}^{0}\to n\overline{K}^{0}) 
&=\mathcal{T}(\Lambda_{b}^{0} \to \Xi^{0}K^{0}),\\
\mathcal{T}(\Xi_{b}^{0}\to p^{+}K^{-})&=\mathcal{T}(\Lambda_{b}^{0}\to 
\Sigma^{+}\pi^{-}),\\
\label{T Rel10}\mathcal{T}(\Xi_{b}^{0}\to 
\Xi^{0}K^{0})&=\mathcal{T}(\Lambda_{b}^{0}\to 
n\overline{K}^{0}),
\end{align}
and the same set of relations for the penguin part,
\begin{align}
\label{P Rel1}
\mathcal{P}(\Lambda_{b}^{0}\to \Sigma^{-}K^{+})&=\mathcal{P}(\Xi_{b}^{0}\to 
\Xi^{-}\pi^{+}),\\
\mathcal{P}(\Lambda_{b}^{0}\to p^{+}\pi^{-})&=\mathcal{P}(\Xi_{b}^{0}\to 
\Sigma^{+}K^{-}),\\
\mathcal{P}(\Xi_{b}^{-}\to nK^{-})&=\mathcal{P}(\Xi_{b}^{-}\to 
\Xi^{0}\pi^{-}),\\
\mathcal{P}(\Xi_{b}^{-}\to \Xi^{-}K^{0})&=\mathcal{P}(\Xi_{b}^{-}\to 
\Sigma^{-}\overline{K}^{0}),\\
\mathcal{P}(\Xi_{b}^{0}\to \Xi^{-}K^{+})&=\mathcal{P}(\Lambda_{b}^{0}\to 
\Sigma^{-}\pi^{+}),\\
\mathcal{P}(\Xi_{b}^{0}\to \Sigma^{-}\pi^{+})&=\mathcal{P}(\Lambda_{b}^{0}\to 
\Xi^{-}K^{+}),\\
\mathcal{P}(\Xi_{b}^{0}\to \Sigma^{+}\pi^{-})&=\mathcal{P}(\Lambda_{b}^{0}\to 
p^{+}K^{-}),\\
\mathcal{P}(\Xi_{b}^{0}\to 
n\overline{K}^{0})&=\mathcal{P}(\Lambda_{b}^{0}\to \Xi^{0}K^{0}),\\
\mathcal{P}(\Xi_{b}^{0}\to p^{+}K^{-})&=\mathcal{P}(\Lambda_{b}^{0}\to 
\Sigma^{+}\pi^{-}),\\
\label{P Rel10}
\mathcal{P}(\Xi_{b}^{0}\to \Xi^{0}K^{0})&=\mathcal{P}(\Lambda_{b}^{0}\to 
n\overline{K}^{0}).
\end{align}
There are several triangle relations connecting the $\Delta S=-1$ decays modes;
\begin{widetext}
\begin{eqnarray}
\mathcal{T}(\Lambda_{b}^{0}\to\Sigma^{+}\pi^{-})+\mathcal{T}(\Lambda_{b}^{0}\to\Sigma^{-}\pi^{+})+2\mathcal{T}(\Lambda_{b}^{0}\to\Sigma^{0}\pi^{0})=0,\nonumber\\
\mathcal{T}(\Xi_{b}^{-}\to\Xi^{-}\pi^{0})-\sqrt{3}\mathcal{T}(\Xi_{b}^{-}\to\Xi^{-}\eta_{8})+\sqrt{2}\mathcal{T}(\Xi_{b}^{-}\to\Sigma^{-}\overline{K^{0}})=0,\nonumber\\
\mathcal{T}(\Xi_{b}^{-}\to\Sigma^{0}K^{-})-\sqrt{3}\mathcal{T}(\Xi_{b}^{-}\to\Lambda^{0}K^{-})+\sqrt{2}\mathcal{T}(\Xi_{b}^{-}\to\Xi^{0}\pi^{-})=0,\nonumber\\
\mathcal{T}(\Xi_{b}^{0}\to\Xi^{-}\pi^{+})-\mathcal{T}(\Lambda_{b}^{0}\to\Xi^{-}K^{+})+\mathcal{T}(\Lambda_{b}^{0}\to\Sigma^{-}\pi^{+})=0,\nonumber\\
\mathcal{T}(\Xi_{b}^{0}\to\Sigma^{+}K^{-})-\mathcal{T}(\Lambda_{b}^{0}\to p^{+}K^{-})+\mathcal{T}(\Lambda_{b}^{0}\to\Sigma^{+}\pi^{-})=0,
\end{eqnarray}
as well as the $\Delta S=0$ decay modes; 
\begin{eqnarray}
\mathcal{T}(\Xi_{b}^{-}\to\Sigma^{0}\pi^{-})-\sqrt{3}\mathcal{T}(\Xi_{b}^{-}\to\Lambda^{0}\pi^{-})-\sqrt{2}\mathcal{T}(\Xi_{b}^{-}\to nK^{-})=0,\nonumber\\
\mathcal{T}(\Xi_{b}^{-}\to\Sigma^{-}\pi^{0})-\sqrt{2}\mathcal{T}(\Xi_{b}^{-}\to\Xi^{-}K^{0})-\sqrt{3}\mathcal{T}(\Xi_{b}^{-}\to \Sigma^{-}\eta_{8})=0,\nonumber\\
\mathcal{T}(\Xi_{b}^{0}\to\Sigma^{-}\pi^{+})-\mathcal{T}(\Xi_{b}^{0}\to\Xi^{-}K^{+})-\mathcal{T}(\Lambda_{b}^{0}\to \Sigma^{-}K^{+})=0,\nonumber\\
\mathcal{T}(\Xi_{b}^{0}\to p^{+}K^{-})-\mathcal{T}(\Xi_{b}^{0}\to\Sigma^{+}\pi^{-})+\mathcal{T}(\Lambda_{b}^{0}\to p^{+}\pi^{-})=0.
\end{eqnarray}
\end{widetext}
The simplest amplitude relations for the case of 
$\overline{\mathbf{3}}_{\mathcal{B}_b}\to 
\mathbf{8}_{\mathcal{B}}\otimes\mathbf{1}_{\mathcal{M}}$ involving the $SU(3)$ 
singlet $\eta_{1}$ are indicated,
\begin{eqnarray}
\mathcal{T}(\Xi_{b}^{0}\to\Xi^{0}\eta_{1})=\mathcal{T}(\Lambda_{b}^{0}\to 
n\eta_{1}),\nonumber\\
\mathcal{T}(\Xi_{b}^{-}\to\Xi^{-}\eta_{1})=\mathcal{T}(\Xi_{b}^{-}\to 
\Sigma^{-}\eta_{1}),
\end{eqnarray}
along with triangle relation for $\Delta S=-1$ processes
\begin{eqnarray}
\mathcal{T}(\Lambda_{b}^{0}\to&\Lambda 
\eta_{1})-\frac{1}{\sqrt{3}}\mathcal{T}(\Lambda_{b}^{0}\to\Sigma^{0}\eta_{1})\nonumber\\&-\frac{\sqrt{2}}{\sqrt{3}}\mathcal{T}(\Xi_{b}^{0}\to
 \Xi^{0}\eta_{1})=0\nonumber
\end{eqnarray}  
and for $\Delta S=0$ processes,
\begin{eqnarray}
\mathcal{T}(\Lambda_{b}^{0}\to n 
\eta_{1})+\frac{\sqrt{3}}{\sqrt{2}}\mathcal{T}(\Xi_{b}^{0}\to\Lambda^{0}\eta_{1})\nonumber\\-\frac{1}{\sqrt{2}}\mathcal{T}(\Xi_{b}^{0}\to
 \Sigma^{0}\eta_{1})=0\nonumber
\end{eqnarray} 

While there is no ground state $SU(3)$ singlet $\Lambda$ baryon, there can be 
$l=1$ excited state spin-3/2 $\Lambda^{0*}_{s}$-baryon, for which one can derive 
amplitude relations in the case of $\overline{\mathbf{3}}_{\mathcal{B}_b}\to 
\mathbf{1}_{\mathcal{B}}\otimes\mathbf{8}_{\mathcal{M}}$; \\
\begin{eqnarray}
\mathcal{T}(\Xi_{b}^{0}\to 
\Lambda_{s}^{0*}\overline{K^{0}})=\mathcal{T}(\Lambda_{b}^{0}\to 
\Lambda_{s}^{0*}K^{0})\nonumber\\
\mathcal{T}(\Xi_{b}^{0}\to \Lambda_{s}^{0*}\eta_{8})=\mathcal{T}(\Xi_{b}^{-}\to 
\Lambda_{s}^{0*}K^{-})
\end{eqnarray}
triangle $\Delta S=-1$ relations:
\begin{eqnarray}
\mathcal{T}(\Lambda_{b}^{0}\to 
&\Lambda^{0*}_{s}\pi_{0})-\frac{1}{\sqrt{3}}\mathcal{T}(\Lambda_{b}^{0}\to 
\Lambda^{0*}_{s}\eta_{8})\nonumber\\&+\frac{\sqrt{2}}{\sqrt{3}}\mathcal{T}(\Xi_{b}^{0}\to
 \Lambda^{0*}_{s}\overline{K^{0}})=0,\nonumber
\end{eqnarray}
triangle $\Delta S=0$ relations:
\begin{eqnarray}
-\frac{1}{\sqrt{3}}\mathcal{T}(\Xi_{b}^{0}\to 
&\Lambda^{0*}_{s}\eta_{8})+\mathcal{T}(\Xi_{b}^{0}\to 
\Lambda^{0*}_{s}\pi_{0})\nonumber\\&-\frac{\sqrt{2}}{\sqrt{3}}\mathcal{T}(\Lambda_{b}^{0}\to
 \Lambda^{0*}_{s}K^{0})=0\nonumber
\end{eqnarray}
The same set of relations hold for penguin part of the all the above mentioned 
amplitude relations.\\
Finally, we consider the trivial case of 
$\overline{\mathbf{3}}_{\mathcal{B}_b}\to 
\mathbf{1}_{\mathcal{B}}\otimes\mathbf{1}_{\mathcal{M}}$ where the final state 
baryon and meson are both $SU(3)$ singlets. The only relevant decay, 
$\Lambda_{b}^{0}\to\Lambda_{s}^{*0}\eta_{1}$, satisfying the $SU(3)$ quantum 
numbers involve a single reduced $SU(3)$ amplitude matching with the counting of 
the number of possible independent $SU(3)$ reduced amplitudes. This concludes our 
discussion of all possible $\overline{\mathbf{3}}_{\mathcal{B}_b}\to 
\mathbf{8}_{\mathcal{B}}\otimes\mathbf{8}_{\mathcal{M}}$, 
$\overline{\mathbf{3}}_{\mathcal{B}_b}\to 
\mathbf{8}_{\mathcal{B}}\otimes\mathbf{1}_{\mathcal{M}}$, 
$\overline{\mathbf{3}}_{\mathcal{B}_b}\to 
\mathbf{1}_{\mathcal{B}}\otimes\mathbf{8}_{\mathcal{M}}$, 
$\overline{\mathbf{3}}_{\mathcal{B}_b}\to 
\mathbf{1}_{\mathcal{B}}\otimes\mathbf{1}_{\mathcal{M}}$ decays of $b$-baryons. 
The most general $SU(3)$ relations can also be obtained in this approach by 
starting from the $T$ matrix and expressing the dependent rows as a linear 
combination of the independent ones. We do not list those relations here as 
they are not particularly illuminating. Nevertheless, in the next section where 
$SU(3)$ breaking effects are taken into account, we do consider a couple of 
interesting $SU(3)$ amplitude relations that should hold under some general 
dynamical assumptions.
%It is important to note that gluonic penguin part of the Hamiltonian described 
%in \ref{P g Ham} only has contributions from \textbf{3} of $SU(3)$ whereas the 
%Electroweak penguins contain \textbf{3}, $\overline{\mathbf{6}}$ and 
%\textbf{15}

\subsection{$SU(3)$ breaking effect}
\label{sec:SU3}
While isospin symmetry holds to a good approximation, the $SU(3)$ symmetry of the light quarks is broken by the mass of the $s$ quark ($m_{s}$). To incorporate such $SU(3)$ violating effects on decay amplitudes, one can parametrize the breaking of flavor $SU(3)$ by the following interaction~\cite{Hinchliffe:1995hz,Pirtskhalava:2011va,Xu:2013dta,Muller:2015rna,Egolf:2002nk,Grossman:2013lya,Xu:2013dta}, \begin{align}
\delta \mathcal{H}=\epsilon \, \overline{q}\,\lambda_{8}\,q
\end{align}
where $\lambda_{8}$ is the Gell-Mann matrix that contributes to the $SU(3)$-breaking and the breaking parameter $\epsilon$ depends on $m_{s}$. The $SU(3)$ structure of the 
unbroken Hamiltonian is modified by this term and to the first order in strange 
quark mass, the broken Hamiltonian is made of the following $SU(3)$ 
representations~\cite{Pirtskhalava:2011va}, 
\begin{multline}
(\mathbf{3}\oplus \overline{\mathbf{6}} \oplus 
\mathbf{15})\otimes(\mathbf{1}+\epsilon 
\, \mathbf{8})=(\mathbf{3}\oplus \overline{\mathbf{6}} \oplus 
\mathbf{15})\nonumber\\+\epsilon (\mathbf{3}_{i}\oplus 
\overline{\mathbf{6}}_{i} \oplus \mathbf{15}_{1} \oplus \mathbf{15}_{2} \oplus 
\mathbf{15}_{3}^{1} \nonumber\\ \oplus \mathbf{15}_{3}^{2}\oplus 
\mathbf{15^{'}}\oplus \mathbf{24}\oplus \mathbf{42}),
\end{multline} 
where the subscript $i=1,2,3$ indicates the origin of that representation from $\mathbf{3}$ ,$\overline{\mathbf{6}}$, $ \mathbf{15}$ respectively. The set of reduced $SU(3)$ amplitudes thus gets enlarged and there are less number of relations as a result. The isospin relation,  
\begin{align}
\mathcal{T}(\Lambda_{b}^{0}\to\Sigma^{+}\pi^{-})&
+\mathcal{T}(\Lambda_{b}^{0}\to\Sigma^{-}\pi^{+})\nonumber \\&
+2\mathcal{T}(\Lambda_{b}^{0}\to\Sigma^{0}\pi^{0})=0
\end{align}
continues to hold  even after including the $SU(3)$ breaking effect to the linear 
order.

There are other amplitude relations that can be derived on more general 
grounds. For instance, the isospin symmetry of the unbroken Hamiltonian forbids a $\Delta 
I=2$ and $\Delta I=5/2$ transition. As a consequence, the $SU(3)$-reduced matrix elements 
 $\braket{\mathbf{f} \parallel 
\mathbf{R}_{I=2} \parallel \mathbf{i}}$ and $\braket{\mathbf{f} \parallel 
\mathbf{R}_{I=5/2} \parallel \mathbf{i}}$ must have a vanishing contribution to the decay amplitude
for arbitrary initial and final states. Such $SU(3)$ breaking but isospin 
conserving relations are given below,
\begin{widetext}
\begin{align}
&\frac{\mathcal{T}(\Xi_{b}^{0}\to\Sigma^{0}\bar{K}^{0})}{3}
	+\frac{\mathcal{T}(\Xi_{b}^{0}\to\Sigma^{+}K^{-})}{3\sqrt{2}}
	+\frac{\mathcal{T}(\Xi_{b}^{0}\to\Xi^{0}\pi^{0})}{3}
	+\frac{\mathcal{T}(\Xi_{b}^{0}\to\Xi^{-}\pi^{+})}{3\sqrt{2}}
\nonumber\\&\qquad	
	+\frac{\mathcal{T}(\Xi_{b}^{-}\to\Sigma^{0}K^{-})}{3}
	+\frac{\mathcal{T}(\Xi_{b}^{-}\to \Sigma^{-}\bar{K}^{0})}{3\sqrt{2}}
	+\frac{\mathcal{T}(\Xi_{b}^{-}\to\Xi^{0}\pi^{-})}{3\sqrt{2}}
	+\frac{\mathcal{T}(\Xi_{b}^{-}\to \Xi^{-}\pi^{0})}{3}
\nonumber\\&\qquad\qquad	\qquad
	+\frac{\sqrt{2} \mathcal{T}(\Lambda_{b}^{0}\to\Sigma^{0}\pi^{0})}{3}
	+\frac{\mathcal{T}(\Lambda_{b}^{0}\to\Sigma^{-}\pi^{+})}{3 \sqrt{2}}
	+\frac{\mathcal{T}(\Lambda_{b}^{0}\to\Sigma^{+}\pi^{-})}{3\sqrt{2}}=0,\\   
&\frac{\mathcal{T}(\Xi_{b}^{0}\to\Sigma^{0}\bar{K}^{0})}{\sqrt{6}}
	+\frac{\mathcal{T}(\Xi_{b}^{0}\to\Sigma^{+}K^{-})}{2\sqrt{3}}
	-\frac{\mathcal{T}(\Xi_{b}^{0}\to\Xi^{0}\pi^{0})}{\sqrt{6}}
	-\frac{\mathcal{T}(\Xi_{b}^{0}\to\Xi^{-}\pi^{+})}{2\sqrt{3}}
\nonumber\\&\qquad\qquad
	+\frac{\mathcal{T}(\Xi_{b}^{-}\to\Sigma^{0}K^{-})}{\sqrt{6}}
	+\frac{\mathcal{T}(\Xi_{b}^{-}\to\Sigma^{-}\bar{K}^{0})}{2\sqrt{3}}
	-\frac{\mathcal{T}(\Xi_{b}^{-}\to\Xi^{0}\pi^{-})}{2\sqrt{3}}
	-\frac{\mathcal{T}(\Xi_{b}^{-}\to\Xi^{-}\pi^{0})}{\sqrt{6}}=0. 
% \nonumber\\&\qquad 
\end{align} 
\end{widetext}
The same set of relations hold for the penguin parts as well.

\section{$CP$ relations} 
\label{sec:CP}
The total decay rate for a two body decay of a spin-$1/2$ anti-triplet 
$b$-baryon ($\mathcal{B}_b$) to a spin 0 pseudo-scalar ($\mathcal{M}$) and a 
spin 
$1/2$ baryon ($\mathcal{B}$) has the following form~\cite{He:2015fwa,Samios:1974tw,Brown:1983wd,Donoghue:1985ww,Dunietz:1992ti} 
\begin{eqnarray}
\label{eq:Gamma-SP}
\Gamma(\mathcal{B}_b\to \mathcal{B}\,\mathcal{M})=\frac{\vert 
	\mathbf{p}_{\mathcal{B}}\vert}{8\pi m_{\mathcal{B}_{b}}^{2}}\Big[\vert S \vert^2 +\vert P \vert^2\Big] \nonumber
\end{eqnarray}
where $\vert\mathbf{p}_{\mathcal{B}}\vert$ is the momentum of the final state 
baryon. Since the decay products can be in any one of the two possible relative 
angular momentum states, $l=0$  and $l=1$, the amplitude can also be decomposed 
in terms of $S$-wave and $P$-wave parts. Including the phase space corrections 
the $S$-wave 
and 
$P$-wave amplitudes are expressed with kinematic factors factored 
out~\cite{Samios:1974tw,Pakvasa:1990if,Dunietz:1992ti} as
\begin{eqnarray}
S&=\sqrt{2 
	m_{\mathcal{B}_{b}}(E_{\mathcal{B}}+m_{\mathcal{B}})}\mathcal{A}^{\mathcal{S}}\nonumber\\
P&=\sqrt{2 
	m_{\mathcal{B}_{b}}(E_{\mathcal{B}}-m_{\mathcal{B}})}\mathcal{A}^{\mathcal{P}}
\label{eq:SP}
\end{eqnarray}   
where $\mathcal{A}^{\mathcal{S}}$ and $\mathcal{A}^{\mathcal{S}}$ are the the 
$SU(3)$-reduced amplitudes defined in Eq.~\eqref{SP wave}. The decay rate is then 
expressed as
\begin{equation}
\Gamma=\frac{\vert \mathbf{p_{\mathcal{B}}} \vert}{4 \pi} 
\frac{(E_{\mathcal{B}}+m_{\mathcal{B}})}{m_{\mathcal{B}_{b}}}\Big[\vert 
\mathcal{A}^{\mathcal{S}} \vert^2+ \Big(\frac{\vert \mathbf{p_{\mathcal{B}}} 
	\vert}{E_{\mathcal{B}}+m_{\mathcal{B}}}\Big)^2 \vert 
\mathcal{A}^{\mathcal{P}} 
\vert^2\Big]
\end{equation} 
$A_{CP}$ is defined subsequently as~\cite{Grossman:2013lya},
\begin{align}
A_{CP}=&\frac{\Gamma(\mathcal{B}_b\to 
	\mathcal{B}\,\mathcal{M})-\Gamma(\overline{\mathcal{B}}_b
	\to\overline{\mathcal{B}}\,\overline{\mathcal{M}})}
	{\Gamma(\mathcal{B}_b\to 
	\mathcal{B}\,\mathcal{M})+\Gamma(\overline{\mathcal{B}}_b
	\to\overline{\mathcal{B}}\,\overline{\mathcal{M}})}\nonumber\\ 			    
	=&\frac{\Delta_{CP}(\mathcal{B}_b\to 
\mathcal{B}\,\mathcal{M})}{2\tilde{\Gamma}(\mathcal{B}_b\to 
\mathcal{B}\,\mathcal{M})},\label{eq:ACP}
\end{align}
where,
$$\tilde{\Gamma}(\mathcal{B}_b\to 
\mathcal{B}\,\mathcal{M})=\frac{1}{2}(\Gamma(\mathcal{B}_b\to 
\mathcal{B}\,\mathcal{M})+\Gamma(\overline{\mathcal{B}}_b\to 
\overline{\mathcal{B}}\,\overline{\mathcal{M}})).$$ 
In order to express  $CP$ relation among modes we rely on the identity 
$\text{Im}(V_{ub}V_{ud}^{*}V_{tb}^{*}V_{td})=
-\text{Im}(V_{ub}V_{us}^{*}V_{tb}^{*}V_{ts})=\textbf{J}$, where $\textbf{J}$ is 
the well known Jarlskog invariant.
Notice, Eqs.~\eqref{eq:Gamma-SP} and \eqref{eq:ACP} imply that $A_{CP}$ is the sum 
$CP$ violation in the $S$ and $P$ waves.
We define a quantity 
$\delta_{CP}^{a}=|\mathcal{A}^a|^2-|\bar{\mathcal{A}}^a|^2$, for 
the partial wave $a$, where $a=\{S,\, P\}$ and $\mathcal{A}^a$ are defined in 
Eq.~\eqref{eq:SP} with phase-space factors removed from the respective partial 
waves.
By definition, 
\begin{multline}
\delta_{CP}^{a}(\mathcal{B}_b\to 
\mathcal{B}\,\mathcal{M})=-\,4\textbf{J}\\
\times\text{Im}\Big[\mathcal{A}^{a*}_{\text{T}}(\mathcal{B}_{b}\to\mathcal{B}\,\mathcal{M})\mathcal{A}^{a}_{\text{P}}(\mathcal{B}_{b}\to\mathcal{B}\,\mathcal{M})\Big].
\end{multline}
Based on amplitude relations for the tree and penguin parts obtained in 
Eqs.~\eqref{T Rel1}--\eqref{T Rel10} and Eqs.~\eqref{P Rel1}--\eqref{P Rel10} 
the 
following ten $\delta_{CP}^{a}$ 
relations are obtained,
\begin{align}
\delta_{CP}^{a}(\Lambda_{b}^{0}\to \Sigma^{-}K^{+})=&
-\delta_{CP}^{a}(\Xi_{b}^{0}\to \Xi^{-}\pi^{+}),\nonumber\\
\delta_{CP}^{a}(\Lambda_{b}^{0}\to p^{+}\pi^{-})=&
-\delta_{CP}^{a}(\Xi_{b}^{0}\to \Sigma^{+}K^{-}),\nonumber\\
\delta_{CP}^{a}(\Xi_{b}^{-}\to n K^{-})=&
-\delta_{CP}^{a}(\Xi_{b}^{-}\to \Xi^{0}\pi^{-}),\nonumber\\
\delta_{CP}^{a}(\Xi_{b}^{-}\to \Xi^{-}K^{0})=&
-\delta_{CP}^{a}(\Xi_{b}^{-}\to \Sigma^{-}\overline{K}^{0}),\nonumber\\
\delta_{CP}^{a}(\Xi_{b}^{0}\to \Xi^{-}K^{+})=&
-\delta_{CP}^{a}(\Lambda_{b}^{0}\to \Sigma^{-}\pi^{+}),\nonumber\\
\delta_{CP}^{a}(\Xi_{b}^{0}\to \Sigma^{-}\pi^{+})=&
-\delta_{CP}^{a}(\Lambda_{b}^{0}\to\Xi^{-}K^{+}),\nonumber\\
\delta_{CP}^{a}(\Xi_{b}^{0}\to \Sigma^{+}\pi^{-})=&
-\delta_{CP}^{a}(\Lambda_{b}^{0}\to p^{+}K^{-}),\nonumber\\
\delta_{CP}^{a}(\Xi_{b}^{0}\to n \overline{K}^{0})=& 
-\delta_{CP}^{a}(\Lambda_{b}^{0}\to \Xi^{0}K^{0}),\nonumber\\
\delta_{CP}^{a}(\Xi_{b}^{0}\to p^{+}K^{-})=&
-\delta_{CP}^{a}(\Lambda_{b}^{0}\to\Sigma^{+}\pi^{-}),\nonumber\\
\delta_{CP}^{a}(\Xi_{b}^{0}\to \Xi^{0}K^{0})=&
-\delta_{CP}^{a}(\Lambda_{b}^{0}\to n \overline{K}^{0}),
\end{align}
for both $a=S$ and $a=P$.
Finally we obtain $A_{CP}$ relations using,
\begin{multline}
A_{CP}(\mathcal{B}_b\to 
\mathcal{B}\,\mathcal{M})\\=\frac{\tau_{\mathcal{B}_b}}{\mathcal{BR}(\mathcal{B}_b\to
	\mathcal{B}\,\mathcal{M})}\Delta_{CP}(\mathcal{B}_b\to 
\mathcal{B}\,\mathcal{M}),
\end{multline}
where, $\tau_{\mathcal{B}}$ is the lifetime of the beauty-baryon. The relation 
between $A_{CP}$ and $\delta_{CP}$ is,
\begin{equation}
\Delta_{CP}=\frac{\vert \mathbf{p_{\mathcal{B}}} \vert}{4 \pi} 
\frac{(E_{\mathcal{B}}+m_{\mathcal{B}})}{m_{\mathcal{B}_{b}}}
%\frac{\vert \mathbf{p_{\mathcal{B}}} \vert}{4 \pi 
%m_{\mathcal{B}_{b}}}(E_{\mathcal{B}}+m_{\mathcal{B}})
\Big[\delta_{CP}^{S}+ 
\Big(\frac{\vert \mathbf{p_{\mathcal{B}}} 
\vert}{E_{\mathcal{B}}+m_{\mathcal{B}}}\Big)^2 \delta_{CP}^{P}\Big]
\end{equation} 
Since, $\Delta_{CP}$ depends on the masses of the initial and final baryons as 
well as the final state meson \cite{Savage:1989qr,Dunietz:1992ti}, some approximation is needed to obtain $A_{CP}$ 
relations between various modes.
Ignoring $\mathbf{p_{\mathcal{B}}}$ and $m_{\mathcal{B}}$ differences between 
the various modes, $CP$ violation relations between various modes can be 
experimentally verified using the relation, 
\begin{equation}
\frac{A_{CP}(\mathcal{B}_{bi}\to\,\mathcal{B}_{j}\mathcal{M}_{k})}
{A_{CP}(\mathcal{B}_{bl}\to \mathcal{B}_{m}\mathcal{M}_{n})}\!
\simeq-\frac{\tau_{\mathcal{B}_{bi}}}
{\tau_{\mathcal{B}_{bl}}}
\frac{\mathcal{BR}(\mathcal{B}_{bl}\to\mathcal{B}_{m}\mathcal{M}_{n})} 
{\mathcal{BR}(\mathcal{B}_{bi}\to\mathcal{B}_{j}\mathcal{M}_{k})},
\end{equation}
where $i$, $j$, $k$ and $l$, $m$, $n$ are indices corresponding to the various 
baryons belonging to the above mentioned $\delta_{CP}$ relations. There is a 
further simplification in case $i=l$, resulting in
\begin{eqnarray}
\frac{A_{CP}(\mathcal{B}_{bi}\to \mathcal{B}_{j}\,\mathcal{M}_{k})}
{A_{CP}(\mathcal{B}_{bl} \to \mathcal{B}_{m}\,\mathcal{M}_{n})}
\simeq-
\frac{\mathcal{BR}(\mathcal{B}_{bl}\to\mathcal{B}_{m}\mathcal{M}_{n})} 
{\mathcal{BR}(\mathcal{B}_{bi}\to\mathcal{B}_{j}\mathcal{M}_{k})},
\end{eqnarray}
where the uncertainties due to lifetime measurement also cancel out~\cite{He:2015fwa}. 
Alternatively, if the longitudinal polarization of the daughter baryon can be 
measured from an angular distribution study of the final states, one can 
estimate the relative strength of the $P$-wave contribution~\cite{Pakvasa:1990if,Dunietz:1992ti} in the total decay 
width. The longitudinal polarization of the daughter baryon is given by, 
\begin{eqnarray}
\alpha=\frac{2\text{Re}(\mathcal{A}^{\mathcal{S}*}\mathcal{A}^{\mathcal{P}})\vert
 \mathbf{p_{\mathcal{B}}} \vert/E_{\mathcal{B}}+m_{\mathcal{B}}}{\vert 
\mathcal{A}^{\mathcal{S}}\vert^{2}+\vert 
\mathcal{A}^{\mathcal{P}}\vert^{2}(\vert \mathbf{p_{\mathcal{B}}} 
\vert/E_{\mathcal{B}}+m_{\mathcal{B}})^{2}}
\end{eqnarray}
The $P$-wave contribution can now be systematically taken into account 
resulting in a more reliable prediction for $A_{CP}$ relations. These relations 
serve as an important test of flavor $SU(3)$ symmetry in  beauty-baryon 
non-leptonic  decays and one can compare these findings with the analogous 
decays of bottom mesons to have a better understanding of the $SU(3)$ flavor 
symmetry breaking pattern. 

%\tcr{(?)This makes it easy to handle the 
%complications due to differing phase-space factors of the $S$ and $P$ wave.}

\section{Conclusions} \label{sec:conclusion} 

We consider a  general framework for hadronic beauty-baryon decays into octet or
singlet of light baryon and a pseudoscalar meson, based on $SU(3)$ decomposition
of the decay amplitudes. We show that in the most general case, the 44 distinct
decay modes require 44 independent reduced $SU(3)$ amplitudes to describe all
possible $\Delta S=-1$ and $\Delta S=0$ processes.  In practice, the dimension-6
effective Hamiltonian that mediates such non-leptonic decays of bottom baryons
predicts only 10 independent reduced $SU(3)$ amplitudes. As a consequence there
must exist relations between the decay amplitudes. We explicitly derive several
sum rules relations between decay amplitudes as well as relations between $CP$ 
asymmetries. Moreover, we systematically study the $SU(3)$-breaking effects in 
the decay amplitudes  at leading order in the $SU(3)$ breaking parameter. We 
further identify an amplitude relation that survives even when the $SU(3)$ 
flavor symmetry is no longer exact. 

\section*{Acknowledgment}
It is a pleasure to thank Benjamin Grinstein for valuable discussions. NGD 
thanks Institute of Mathematical Sciences for hospitality where part of the 
work was done. RS thanks the Institute of Theoretical Science, University of Oregon, Eugene, Oregon for hospitality where part of the work was done. This research was supported in part by Perimeter Institute for 
Theoretical Physics. Research at Perimeter Institute is supported by the 
Government of Canada through the Department of Innovation, Science and Economic 
Development and by the Province of Ontario through the Ministry of Research, 
Innovation and Science.

\appendix
%\newpage
\begin{widetext}
\section{$SU(3)$ isoscalar factors}
\label{sec:App0}
The isoscalar factors depend of the identity of the representations, and on the 
hypercharges and isospins of the isomultiplets that are coupled. Let us denote 
them by the following notation;
\begin{equation}
\begin{pmatrix}
\mathbf{r}       & \mathbf{r^{'}}  & \mathbf{R} \\
\noalign{\smallskip}    
(y,i,i_{3}) & (y^{'},i^{'},i^{'}_{3}) 	& 
(Y,I,I_{3}) 
\end{pmatrix}.
\end{equation}

The $SU(3)$ Clebsch-Gordan coefficients are found as products of isoscalar factors
and $SU(2)$ Clebsch-Gordan coefficients: 
\begin{align*}
\langle \mathbf{R},Y,I,I_{3}\vert 
\mathbf{r},y,i,i_{3},\mathbf{r^{'}},y^{'},i^{'},i^{'}_{3}  
\rangle=\begin{pmatrix}
\mathbf{r}       & \mathbf{r^{'}}  & \mathbf{R} \\
\noalign{\smallskip}    
(y,i,i_{3}) & (y^{'},i^{'},i^{'}_{3}) 	& 
(Y,I,I_{3}) 
\end{pmatrix} \langle I,I_{3} \vert i,i_{3},i^{'},i^{'}_{3} \rangle
\end{align*}
where $\langle I,I_{3} \vert i,i_{3},i^{'},i^{'}_{3} \rangle$ are the $SU(2)$ CG 
coefficients. The order in which the $SU(3)$ representations are coupled is 
$\mathbf{r}\otimes \mathbf{r^{'}}\to\mathbf{R}$. The two symmetry relations 
involving the $SU(3)$ isoscalar factors are as follows;

A) If the order in which the representations are coupled is reversed (i.e. 
$\mathbf{r^{'}}\otimes \mathbf{r}\to\mathbf{R}$) then the isoscalar factors 
pick up a phase factor;
\begin{align}
\label{A2}
\begin{pmatrix}
\mathbf{r^{'}}       & \mathbf{r}  & \mathbf{R} \\
\noalign{\smallskip}    
(y^{'},i^{'},i^{'}_{3}) & (y,i,i_{3})  	& 
(Y,I,I_{3}) 
\end{pmatrix}
=(-1)^{I-i-i^{'}}\mathbf{\xi}(\mathbf{R};\mathbf{r},\mathbf{r^{'}})\begin{pmatrix}
\mathbf{r}       & \mathbf{r^{'}}  & \mathbf{R} \\
\noalign{\smallskip}    
(y,i,i_{3}) & (y^{'},i^{'},i^{'}_{3}) 	& 
(Y,I,I_{3}) 
\end{pmatrix}.
\end{align}
Here $\mathbf{\xi}(\mathbf{R};\mathbf{r},\mathbf{r^{'}})$ is the phase factor 
\cite{Kaeding:1995vq} that depends only on the identity element of 
$\mathbf{r}$, $\mathbf{r^{'}}$ and $\mathbf{R}$ and the phase convention chosen 
for the highest weight state. \\

B) Conjugation operation on all three representations also give rise to a phase 
factor;
\begin{align}
\label{A3}
\begin{pmatrix}
\overline{\mathbf{r}}       & \overline{\mathbf{r^{'}}}  & 
\overline{\mathbf{R}} \\
\noalign{\smallskip}    
(y,i,i_{3}) & (y^{'},i^{'},i^{'}_{3}) 	& 
(Y,I,I_{3}) 
\end{pmatrix}
=(-1)^{I-i-i^{'}}\mathbf{\zeta}(\mathbf{R};\mathbf{r},\mathbf{r^{'}})\begin{pmatrix}
\mathbf{r}       & \mathbf{r^{'}}  & \mathbf{R} \\
\noalign{\smallskip}    
(-y,i,-i_{3}) & (-y^{'},i^{'},-i^{'}_{3}) 	& 
(-Y,I,-I_{3}) 
\end{pmatrix}.
\end{align}
\end{widetext}
Similar to the previous case, 
$\mathbf{\zeta}(\mathbf{R};\mathbf{r},\mathbf{r^{'}})$ is the phase factor 
\cite{Kaeding:1995vq} that depends only on the identity element of 
$\mathbf{r}$, $\mathbf{r^{'}}$ and $\mathbf{R}$ and the phase convention chosen 
for the highest weight state. As a corollary of Eqs.~\eqref{A2} and 
\eqref{A3},\begin{eqnarray}
\mathbf{\xi}(\overline{\mathbf{R}};\overline{\mathbf{r}},\overline{\mathbf{r}^{'}})=\mathbf{\xi}(\mathbf{R};\mathbf{r},\mathbf{r^{'}}),\\
\mathbf{\zeta}(\mathbf{R};\mathbf{r^{'}},\mathbf{r})=\mathbf{\zeta}(\mathbf{R};\mathbf{r},\mathbf{r^{'}})
\end{eqnarray}  
Following \cite{Kaeding:1995vq}, we have adopted the Condon-Shortley and de 
Swart phase convention \cite{deSwart:1963pdg,Shortley} that requires 
eigenvalues of the isospin(I) as well as $V$ spin raising and lowering 
operators are real and positive. An additional requirement on the highest 
weight state is 
Clebsch-Gordan coefficient between these three states be real and positive, 
i.e. \begin{eqnarray}
\langle \mathbf{R},Y_{h},I_{h},I_{h3}\vert 
\mathbf{r},y_{h},i_{h},i_{h3},\mathbf{r^{'}},y^{'}_{h},i^{'}_{h},i^{'}_{h3}  
\rangle > 0 \nonumber
\end{eqnarray}
These conditions ensure that $SU(3)$ CG coefficients and isoscalar factors are 
all real.
\begin{widetext}
\section{$SU(3)$ decomposition of $\Delta S=0$ processes}
\label{sec:App1}
The $\Delta S=0$ decay amplitudes for 
$\overline{\textbf{3}}_{\mathcal{B}_{b}}\to \textbf{8}_{\mathcal{B}}\, 
\textbf{8}_{\mathcal{M}}$  are $SU(3)$ decomposed in the following way;
\begin{equation}
\centering
\rotatebox{90}{
$
\resizebox{\vsize}{!}{%
$
\setcounter{MaxMatrixCols}{22}
\begin{pmatrix}
\mathcal{A}(\Lambda_{b}^{0}\to n\eta_{8})\\ 
\mathcal{A}(\Lambda_{b}^{0}\to \Sigma^{-}K^{+})\\ 
\mathcal{A}(\Lambda_{b}^{0}\to p^{+}\pi^{-})\\ 
\mathcal{A}(\Lambda_{b}^{0}\to n\pi^{0})\\ 
\mathcal{A}(\Lambda_{b}^{0}\to \Lambda^{0}K^{0})\\ 
\mathcal{A}(\Lambda_{b}^{0}\to \Sigma^{0}K^{0})\\ 
\mathcal{A}(\Xi_{b}^{-}\to nK^{-})\\
\mathcal{A}(\Xi_{b}^{-}\to \Sigma^{-}\eta_{8})\\ 
\mathcal{A}(\Xi_{b}^{-}\to \Lambda^{0}\pi^{-})\\
\mathcal{A}(\Xi_{b}^{-}\Xi^{-}K^{0})\\ 
\mathcal{A}(\Xi_{b}^{-}\to \Sigma^{0}\pi^{-})\\ 
\mathcal{A}(\Xi_{b}^{-}\to \Sigma^{-}\pi^{0})\\
\mathcal{A}(\Xi_{b}^{0}\Xi^{-}K^{+})\\ 
\mathcal{A}(\Xi_{b}^{0}\Sigma^{-}\pi^{+})\\ 
\mathcal{A}(\Xi_{b}^{0}\to \Sigma^{+}\pi^{-})\\ 
\mathcal{A}(\Xi_{b}^{0}\to n\overline{K^{0}})\\
\mathcal{A}(\Xi_{b}^{0}\to \Sigma^{0}\eta_{8})\\
\mathcal{A}(\Xi_{b}^{0}\to \Sigma^{0}\pi^{0})\\ 
\mathcal{A}(\Xi_{b}^{0}\to p^{+}K^{-})\\
\mathcal{A}(\Xi_{b}^{0}\Xi^{0}K^{0})\\
\mathcal{A}(\Xi_{b}^{0}\to \Lambda^{0}\eta_8)\\ 
\mathcal{A}(\Xi_{b}^{0}\to \Lambda^{0}\pi^{0})
\end{pmatrix}
=\begin{pmatrix}
 \frac{1}{\sqrt{5}} & \frac{1}{2 \sqrt{5}} & \frac{1}{\sqrt{5}} & \frac{\sqrt{\frac{3}{5}}}{2} & \frac{1}{\sqrt{10}} &
   -\frac{\sqrt{\frac{3}{10}}}{4} & -\frac{\sqrt{\frac{3}{10}}}{4} & -\frac{1}{4 \sqrt{5}} & \frac{\sqrt{\frac{3}{2}}}{4} &
   \frac{\sqrt{\frac{3}{2}}}{4} & \frac{1}{4} & 0 & 0 & 0 & 0 & 0 & 0 & 0 & 0 & 0 & 0 & 0 \\
 \frac{\sqrt{\frac{2}{15}}}{3} & \frac{1}{3 \sqrt{30}} & \frac{\sqrt{\frac{2}{15}}}{3} & \frac{1}{\sqrt{10}} & \frac{1}{\sqrt{15}} &
   \frac{3}{4 \sqrt{5}} & \frac{3}{4 \sqrt{5}} & \frac{\sqrt{\frac{3}{10}}}{2} & -\frac{1}{4} & -\frac{1}{4} & -\frac{1}{2 \sqrt{6}} &
   \frac{1}{2 \sqrt{2}} & \frac{1}{2 \sqrt{6}} & \frac{\sqrt{\frac{5}{3}}}{6} & \frac{\sqrt{\frac{2}{3}}}{3} &
   \frac{\sqrt{\frac{5}{3}}}{6} & 0 & 0 & 0 & 0 & 0 & 0 \\
 \frac{\sqrt{\frac{2}{15}}}{3} & \frac{1}{3 \sqrt{30}} & \frac{\sqrt{\frac{2}{15}}}{3} & -\frac{1}{\sqrt{10}} & -\frac{1}{\sqrt{15}} &
   \frac{3}{4 \sqrt{5}} & \frac{3}{4 \sqrt{5}} & \frac{\sqrt{\frac{3}{10}}}{2} & \frac{1}{4} & \frac{1}{4} & \frac{1}{2 \sqrt{6}} &
   -\frac{1}{2 \sqrt{2}} & -\frac{1}{2 \sqrt{6}} & \frac{\sqrt{\frac{5}{3}}}{6} & \frac{\sqrt{\frac{2}{3}}}{3} &
   \frac{\sqrt{\frac{5}{3}}}{6} & 0 & 0 & 0 & 0 & 0 & 0 \\
 -\frac{1}{3 \sqrt{15}} & -\frac{1}{6 \sqrt{15}} & -\frac{1}{3 \sqrt{15}} & \frac{1}{2 \sqrt{5}} & \frac{1}{\sqrt{30}} & -\frac{3}{4
   \sqrt{10}} & -\frac{3}{4 \sqrt{10}} & -\frac{\sqrt{\frac{3}{5}}}{4} & -\frac{1}{4 \sqrt{2}} & -\frac{1}{4 \sqrt{2}} & -\frac{1}{4
   \sqrt{3}} & -\frac{1}{2} & -\frac{1}{2 \sqrt{3}} & \frac{\sqrt{\frac{5}{6}}}{3} & \frac{2}{3 \sqrt{3}} & \frac{\sqrt{\frac{5}{6}}}{3}
   & 0 & 0 & 0 & 0 & 0 & 0 \\
 \frac{1}{\sqrt{5}} & \frac{1}{2 \sqrt{5}} & \frac{1}{\sqrt{5}} & -\frac{\sqrt{\frac{3}{5}}}{2} & -\frac{1}{\sqrt{10}} &
   -\frac{\sqrt{\frac{3}{10}}}{4} & -\frac{\sqrt{\frac{3}{10}}}{4} & -\frac{1}{4 \sqrt{5}} & -\frac{\sqrt{\frac{3}{2}}}{4} &
   -\frac{\sqrt{\frac{3}{2}}}{4} & -\frac{1}{4} & 0 & 0 & 0 & 0 & 0 & 0 & 0 & 0 & 0 & 0 & 0 \\
 -\frac{1}{3 \sqrt{15}} & -\frac{1}{6 \sqrt{15}} & -\frac{1}{3 \sqrt{15}} & -\frac{1}{2 \sqrt{5}} & -\frac{1}{\sqrt{30}} & -\frac{3}{4
   \sqrt{10}} & -\frac{3}{4 \sqrt{10}} & -\frac{\sqrt{\frac{3}{5}}}{4} & \frac{1}{4 \sqrt{2}} & \frac{1}{4 \sqrt{2}} & \frac{1}{4
   \sqrt{3}} & \frac{1}{2} & \frac{1}{2 \sqrt{3}} & \frac{\sqrt{\frac{5}{6}}}{3} & \frac{2}{3 \sqrt{3}} & \frac{\sqrt{\frac{5}{6}}}{3} &
   0 & 0 & 0 & 0 & 0 & 0 \\
 \frac{2 \sqrt{\frac{2}{15}}}{3} & -\frac{2 \sqrt{\frac{2}{15}}}{3} & -\frac{\sqrt{\frac{2}{15}}}{3} & -\frac{\sqrt{\frac{2}{5}}}{3} &
   \frac{1}{\sqrt{15}} & -\frac{3}{4 \sqrt{5}} & \frac{1}{4 \sqrt{5}} & \frac{\sqrt{\frac{3}{10}}}{2} & \frac{1}{4} & -\frac{1}{12} &
   -\frac{1}{2 \sqrt{6}} & -\frac{1}{6 \sqrt{2}} & \frac{1}{2 \sqrt{6}} & \frac{1}{6 \sqrt{15}} & \frac{1}{3 \sqrt{6}} &
   -\frac{\sqrt{\frac{5}{3}}}{6} & \frac{1}{3} & -\frac{1}{3 \sqrt{2}} & \frac{1}{\sqrt{10}} & -\frac{1}{3 \sqrt{2}} & 0 & 0 \\
 \frac{2}{3 \sqrt{5}} & -\frac{2}{3 \sqrt{5}} & -\frac{1}{3 \sqrt{5}} & -\frac{1}{\sqrt{15}} & \frac{1}{\sqrt{10}} &
   \frac{\sqrt{\frac{3}{10}}}{2} & -\frac{1}{2 \sqrt{30}} & -\frac{1}{2 \sqrt{5}} & 0 & 0 & 0 & \frac{1}{4 \sqrt{3}} & -\frac{1}{4} &
   \frac{1}{6 \sqrt{10}} & \frac{1}{6} & -\frac{\sqrt{\frac{5}{2}}}{6} & -\frac{1}{\sqrt{6}} & -\frac{1}{2 \sqrt{3}} &
   -\frac{1}{\sqrt{15}} & 0 & 0 & 0 \\
 \frac{2}{3 \sqrt{5}} & -\frac{2}{3 \sqrt{5}} & -\frac{1}{3 \sqrt{5}} & \frac{1}{\sqrt{15}} & -\frac{1}{\sqrt{10}} &
   \frac{\sqrt{\frac{3}{10}}}{2} & -\frac{1}{2 \sqrt{30}} & -\frac{1}{2 \sqrt{5}} & 0 & 0 & 0 & -\frac{1}{4 \sqrt{3}} & \frac{1}{4} &
   \frac{1}{6 \sqrt{10}} & \frac{1}{6} & -\frac{\sqrt{\frac{5}{2}}}{6} & \frac{1}{\sqrt{6}} & \frac{1}{2 \sqrt{3}} &
   -\frac{1}{\sqrt{15}} & 0 & 0 & 0 \\
 \frac{2 \sqrt{\frac{2}{15}}}{3} & -\frac{2 \sqrt{\frac{2}{15}}}{3} & -\frac{\sqrt{\frac{2}{15}}}{3} & \frac{\sqrt{\frac{2}{5}}}{3} &
   -\frac{1}{\sqrt{15}} & -\frac{3}{4 \sqrt{5}} & \frac{1}{4 \sqrt{5}} & \frac{\sqrt{\frac{3}{10}}}{2} & -\frac{1}{4} & \frac{1}{12} &
   \frac{1}{2 \sqrt{6}} & \frac{1}{6 \sqrt{2}} & -\frac{1}{2 \sqrt{6}} & \frac{1}{6 \sqrt{15}} & \frac{1}{3 \sqrt{6}} &
   -\frac{\sqrt{\frac{5}{3}}}{6} & -\frac{1}{3} & \frac{1}{3 \sqrt{2}} & \frac{1}{\sqrt{10}} & \frac{1}{3 \sqrt{2}} & 0 & 0 \\
 0 & 0 & 0 & \frac{1}{3 \sqrt{5}} & -\frac{1}{\sqrt{30}} & 0 & 0 & 0 & \frac{1}{2 \sqrt{2}} & -\frac{1}{6 \sqrt{2}} & -\frac{1}{2
   \sqrt{3}} & \frac{1}{12} & -\frac{1}{4 \sqrt{3}} & \frac{\sqrt{\frac{5}{6}}}{2} & -\frac{1}{2 \sqrt{3}} & -\frac{1}{2 \sqrt{30}} &
   -\frac{1}{3 \sqrt{2}} & \frac{1}{6} & 0 & -\frac{1}{3} & -\frac{1}{\sqrt{5}} & 0 \\
 0 & 0 & 0 & -\frac{1}{3 \sqrt{5}} & \frac{1}{\sqrt{30}} & 0 & 0 & 0 & -\frac{1}{2 \sqrt{2}} & \frac{1}{6 \sqrt{2}} & \frac{1}{2
   \sqrt{3}} & -\frac{1}{12} & \frac{1}{4 \sqrt{3}} & \frac{\sqrt{\frac{5}{6}}}{2} & -\frac{1}{2 \sqrt{3}} & -\frac{1}{2 \sqrt{30}} &
   \frac{1}{3 \sqrt{2}} & -\frac{1}{6} & 0 & \frac{1}{3} & -\frac{1}{\sqrt{5}} & 0 \\
 \frac{1}{6 \sqrt{30}} & -\frac{\sqrt{\frac{5}{6}}}{3} & \frac{\sqrt{\frac{2}{15}}}{3} & \frac{1}{3 \sqrt{10}} & -\frac{1}{2 \sqrt{15}}
   & -\frac{1}{2 \sqrt{5}} & \frac{1}{2 \sqrt{5}} & 0 & 0 & -\frac{1}{3} & \frac{1}{\sqrt{6}} & -\frac{1}{6 \sqrt{2}} & \frac{1}{2
   \sqrt{6}} & -\frac{1}{6 \sqrt{15}} & -\frac{1}{3 \sqrt{6}} & \frac{\sqrt{\frac{5}{3}}}{6} & -\frac{1}{6} & -\frac{1}{3 \sqrt{2}} &
   -\frac{1}{\sqrt{10}} & -\frac{1}{3 \sqrt{2}} & 0 & \frac{1}{2 \sqrt{2}} \\
 \frac{1}{6 \sqrt{30}} & \frac{1}{3 \sqrt{30}} & -\frac{1}{3 \sqrt{30}} & -\frac{1}{3 \sqrt{10}} & \frac{1}{2 \sqrt{15}} & \frac{1}{4
   \sqrt{5}} & -\frac{3}{4 \sqrt{5}} & \frac{\sqrt{\frac{3}{10}}}{2} & -\frac{1}{4} & \frac{1}{12} & \frac{1}{2 \sqrt{6}} & \frac{1}{6
   \sqrt{2}} & -\frac{1}{2 \sqrt{6}} & \frac{\sqrt{\frac{5}{3}}}{6} & -\frac{1}{3 \sqrt{6}} & -\frac{1}{6 \sqrt{15}} & \frac{1}{6} &
   \frac{1}{3 \sqrt{2}} & 0 & -\frac{\sqrt{2}}{3} & \frac{1}{\sqrt{10}} & \frac{1}{2 \sqrt{2}} \\
 \frac{1}{6 \sqrt{30}} & \frac{1}{3 \sqrt{30}} & -\frac{1}{3 \sqrt{30}} & \frac{1}{3 \sqrt{10}} & -\frac{1}{2 \sqrt{15}} & \frac{1}{4
   \sqrt{5}} & -\frac{3}{4 \sqrt{5}} & \frac{\sqrt{\frac{3}{10}}}{2} & \frac{1}{4} & -\frac{1}{12} & -\frac{1}{2 \sqrt{6}} & -\frac{1}{6
   \sqrt{2}} & \frac{1}{2 \sqrt{6}} & \frac{\sqrt{\frac{5}{3}}}{6} & -\frac{1}{3 \sqrt{6}} & -\frac{1}{6 \sqrt{15}} & -\frac{1}{6} &
   -\frac{1}{3 \sqrt{2}} & 0 & \frac{\sqrt{2}}{3} & \frac{1}{\sqrt{10}} & \frac{1}{2 \sqrt{2}} \\
 \frac{7}{6 \sqrt{30}} & \frac{1}{3 \sqrt{30}} & -\frac{2 \sqrt{\frac{2}{15}}}{3} & -\frac{1}{3 \sqrt{10}} & \frac{1}{2 \sqrt{15}} &
   -\frac{1}{4 \sqrt{5}} & -\frac{1}{4 \sqrt{5}} & \frac{\sqrt{\frac{3}{10}}}{2} & \frac{1}{4} & -\frac{5}{12} & \frac{1}{2 \sqrt{6}} &
   \frac{1}{6 \sqrt{2}} & -\frac{1}{2 \sqrt{6}} & -\frac{1}{6 \sqrt{15}} & -\frac{1}{3 \sqrt{6}} & \frac{\sqrt{\frac{5}{3}}}{6} &
   \frac{1}{6} & \frac{1}{3 \sqrt{2}} & -\frac{1}{\sqrt{10}} & \frac{1}{3 \sqrt{2}} & 0 & -\frac{1}{2 \sqrt{2}} \\
 \frac{\sqrt{\frac{2}{5}}}{3} & -\frac{\sqrt{\frac{2}{5}}}{3} & -\frac{1}{3 \sqrt{10}} & -\frac{1}{\sqrt{30}} & \frac{1}{2 \sqrt{5}} &
   \frac{\sqrt{\frac{3}{5}}}{4} & -\frac{1}{4 \sqrt{15}} & -\frac{1}{2 \sqrt{10}} & 0 & 0 & 0 & -\frac{1}{2 \sqrt{6}} & \frac{1}{2
   \sqrt{2}} & -\frac{1}{6 \sqrt{5}} & -\frac{1}{3 \sqrt{2}} & \frac{\sqrt{5}}{6} & -\frac{1}{2 \sqrt{3}} & \frac{1}{\sqrt{6}} &
   \sqrt{\frac{2}{15}} & 0 & 0 & 0 \\
 -\frac{1}{6 \sqrt{30}} & -\frac{1}{3 \sqrt{30}} & \frac{1}{3 \sqrt{30}} & 0 & 0 & -\frac{1}{4 \sqrt{5}} & \frac{3}{4 \sqrt{5}} &
   -\frac{\sqrt{\frac{3}{10}}}{2} & 0 & 0 & 0 & 0 & 0 & \frac{\sqrt{\frac{5}{3}}}{3} & -\frac{\sqrt{\frac{2}{3}}}{3} & -\frac{1}{3
   \sqrt{15}} & 0 & 0 & 0 & 0 & \sqrt{\frac{2}{5}} & -\frac{1}{2 \sqrt{2}} \\
 \frac{1}{6 \sqrt{30}} & -\frac{\sqrt{\frac{5}{6}}}{3} & \frac{\sqrt{\frac{2}{15}}}{3} & -\frac{1}{3 \sqrt{10}} & \frac{1}{2 \sqrt{15}}
   & -\frac{1}{2 \sqrt{5}} & \frac{1}{2 \sqrt{5}} & 0 & 0 & \frac{1}{3} & -\frac{1}{\sqrt{6}} & \frac{1}{6 \sqrt{2}} & -\frac{1}{2
   \sqrt{6}} & -\frac{1}{6 \sqrt{15}} & -\frac{1}{3 \sqrt{6}} & \frac{\sqrt{\frac{5}{3}}}{6} & \frac{1}{6} & \frac{1}{3 \sqrt{2}} &
   -\frac{1}{\sqrt{10}} & \frac{1}{3 \sqrt{2}} & 0 & \frac{1}{2 \sqrt{2}} \\
 \frac{7}{6 \sqrt{30}} & \frac{1}{3 \sqrt{30}} & -\frac{2 \sqrt{\frac{2}{15}}}{3} & \frac{1}{3 \sqrt{10}} & -\frac{1}{2 \sqrt{15}} &
   -\frac{1}{4 \sqrt{5}} & -\frac{1}{4 \sqrt{5}} & \frac{\sqrt{\frac{3}{10}}}{2} & -\frac{1}{4} & \frac{5}{12} & -\frac{1}{2 \sqrt{6}} &
   -\frac{1}{6 \sqrt{2}} & \frac{1}{2 \sqrt{6}} & -\frac{1}{6 \sqrt{15}} & -\frac{1}{3 \sqrt{6}} & \frac{\sqrt{\frac{5}{3}}}{6} &
   -\frac{1}{6} & -\frac{1}{3 \sqrt{2}} & -\frac{1}{\sqrt{10}} & -\frac{1}{3 \sqrt{2}} & 0 & -\frac{1}{2 \sqrt{2}} \\
 -\frac{\sqrt{\frac{3}{10}}}{2} & -\sqrt{\frac{3}{10}} & \sqrt{\frac{3}{10}} & 0 & 0 & \frac{1}{4 \sqrt{5}} & -\frac{3}{4 \sqrt{5}} &
   \frac{\sqrt{\frac{3}{10}}}{2} & 0 & 0 & 0 & 0 & 0 & 0 & 0 & 0 & 0 & 0 & 0 & 0 & 0 & -\frac{1}{2 \sqrt{2}} \\
 \frac{\sqrt{\frac{2}{5}}}{3} & -\frac{\sqrt{\frac{2}{5}}}{3} & -\frac{1}{3 \sqrt{10}} & \frac{1}{\sqrt{30}} & -\frac{1}{2 \sqrt{5}} &
   \frac{\sqrt{\frac{3}{5}}}{4} & -\frac{1}{4 \sqrt{15}} & -\frac{1}{2 \sqrt{10}} & 0 & 0 & 0 & \frac{1}{2 \sqrt{6}} & -\frac{1}{2
   \sqrt{2}} & -\frac{1}{6 \sqrt{5}} & -\frac{1}{3 \sqrt{2}} & \frac{\sqrt{5}}{6} & \frac{1}{2 \sqrt{3}} & -\frac{1}{\sqrt{6}} &
   \sqrt{\frac{2}{15}} & 0 & 0 & 0 \\
\end{pmatrix}
\begin{pmatrix}
\langle 27 \parallel \mathbf{15_{1/2}}\parallel \overline{3}\rangle \\
\langle 27\parallel \mathbf{24_{1/2}}\parallel \overline{3}\rangle\\
\langle 27\parallel \mathbf{42_{1/2}}\parallel \overline{3} \rangle \\ 
\langle \overline{10}\parallel \mathbf{24_{1/2}} \parallel \overline{3}\rangle\\ 
\langle \overline{10}\parallel \mathbf{\overline{6}_{1/2}} \parallel \parallel \overline{3}\rangle \\
\langle 8_{1}\parallel \mathbf{3_{1/2}}\parallel \overline{3}\rangle\\
\langle 8_{1}\parallel \mathbf{15_{1/2}}\parallel \overline{3}\rangle\\ 
\langle 8_{1}\parallel\mathbf{\overline{6}_{1/2}}\parallel \overline{3}\\ 
\langle 8_{2}\parallel \mathbf{3_{1/2}}\parallel \overline{3}\\ 
\langle 8_{2}\parallel \mathbf{15_{1/2}}\parallel \overline{3}\\ 
\langle 8_{2}\parallel \mathbf{\overline{6}_{1/2}}\parallel \overline{3}\\ 
\langle 10 \parallel \mathbf{15_{3/2}}\parallel \overline{3}\\
\langle 10\parallel \mathbf{15^{'}_{3/2}}\parallel \overline{3}\\
\langle 27\parallel \mathbf{15_{3/2}}\parallel \overline{3}\\ 
\langle27\parallel \mathbf{24_{3/2}}\parallel \overline{3}\\
\langle 27\parallel \mathbf{42_{3/2}}\parallel \overline{3}\\
\langle 10 \parallel \mathbf{15_{1/2}}\parallel \overline{3}\\ 
\langle 10 \parallel \mathbf{24_{3/2}}\parallel \overline{3}\\
\langle 8_{1}\parallel \mathbf{15_{3/2}}\parallel \overline{3}\\ 
\langle 8_{2}\parallel \mathbf{15_{3/2}}\parallel \overline{3}\\ 
\langle 27\parallel \mathbf{42_{5/2}}\parallel \overline{3}\\ 
\langle 1\parallel \mathbf{3_{1/2}}\parallel \overline{3}
\end{pmatrix}
$
	}
$
	}
\nonumber	
\end{equation}
The $\Delta S=0$ decay amplitudes for 
$\overline{\textbf{3}}_{\mathcal{B}_{b}}\to \textbf{8}_{\mathcal{B}}\, 
\textbf{1}_{\mathcal{M}}$  are $SU(3)$ decomposed in the following way:
\begin{equation}
\begin{pmatrix}
\mathcal{A}(\Lambda_{b}^{0} \to n\eta_{1})\\
\mathcal{A}(\Xi_{b}^{0} \to \Lambda^{0}\eta_{1}) \\
\mathcal{A}(\Xi_{b}^{0} \to \Sigma^{0}\eta_{1})\\
\mathcal{A}(\Xi_{b}^{-} \to \Sigma^{-}\eta_{1})\\
\end{pmatrix}=\begin{pmatrix}
\frac{\sqrt{\frac{3}{2}}}{2} & \frac{1}{2} & \frac{\sqrt{\frac{3}{2}}}{2} & 0 
\\
-\frac{1}{4} & -\frac{\sqrt{\frac{3}{2}}}{2} & \frac{3}{4} & 0 \\
\frac{\sqrt{3}}{4} & -\frac{1}{2 \sqrt{2}} & -\frac{1}{4 \sqrt{3}} & 
\sqrt{\frac{2}{3}} \\
\frac{\sqrt{\frac{3}{2}}}{2} & -\frac{1}{2} & -\frac{1}{2 \sqrt{6}} & 
-\frac{1}{\sqrt{3}} \\
\end{pmatrix}
\begin{pmatrix}
\langle \mathbf{8} \parallel \mathbf{3}_{1/2} \parallel \overline{3} \rangle \\
\langle \mathbf{8} \parallel \overline{\mathbf{6}}_{1/2} \parallel \overline{3} 
\rangle \\
\langle \mathbf{8} \parallel \mathbf{15}_{1/2} \parallel \overline{3} \rangle \\
\langle \mathbf{8} \parallel \mathbf{15}_{3/2} \parallel \overline{3} \rangle
\end{pmatrix}
\end{equation}
The $\Delta S=0$ decay amplitudes for 
$\overline{\textbf{3}}_{\mathcal{B}_{b}}\to \textbf{1}_{\mathcal{B}}\, 
\textbf{8}_{\mathcal{M}}$  are $SU(3)$ decomposed in the following way;
\begin{equation}
\begin{pmatrix}
\mathcal{A}(\Xi_{b}^{0}\to \Lambda_{s}^{0*}\eta_{8})\\
\mathcal{A}(\Xi_{b}^{0}\to \Lambda_{s}^{0*}\pi_{0})\\
\mathcal{A}(\Xi_{b}^{-}\to \Lambda_{s}^{0*}\pi^{-})\\
\mathcal{A}(\Lambda_{b}^{0}\to \Lambda_{s}^{0*}K^{0})\\
\end{pmatrix}
=\begin{pmatrix}
-\frac{1}{4} & -\frac{\sqrt{\frac{3}{2}}}{2} & \frac{3}{4} & 0 \\
\frac{\sqrt{3}}{4} & -\frac{1}{2 \sqrt{2}} & -\frac{1}{4 \sqrt{3}} & 
\sqrt{\frac{2}{3}} \\
\frac{\sqrt{\frac{3}{2}}}{2} & -\frac{1}{2} & -\frac{1}{2 \sqrt{6}} & 
-\frac{1}{\sqrt{3}} \\
\frac{\sqrt{\frac{3}{2}}}{2} & \frac{1}{2} & \frac{\sqrt{\frac{3}{2}}}{2} & 0 
\\
\end{pmatrix}
\begin{pmatrix}
\langle \mathbf{8} \parallel \mathbf{3}_{1/2} \parallel \overline{3} \rangle \\
\langle \mathbf{8} \parallel \overline{\mathbf{6}}_{1/2} \parallel \overline{3} 
\rangle \\
\langle \mathbf{8} \parallel \mathbf{15}_{1/2} \parallel \overline{3} \rangle \\
\langle \mathbf{8} \parallel \mathbf{15}_{3/2} \parallel \overline{3} \rangle
\end{pmatrix}
\end{equation}

\section{$SU(3)$ decomposition of $\Delta S=-1$ processes}
\label{sec:App2}
The $\Delta S=-1$ decay amplitudes for $\overline{\textbf{3}}_{\mathcal{B}_{b}}\to \textbf{8}_{\mathcal{B}}\, 
\textbf{1}_{\mathcal{M}}$   are $SU(3)$ decomposed in the following way;
\begin{equation}
\begin{pmatrix}
 \mathcal{A}(\Lambda_{b}^{0} \to \Lambda^{0}\eta_{1})\\
 \mathcal{A}(\Lambda_{b}^{0} \to \Sigma^{0}\eta_{1}) \\
 \mathcal{A}(\Xi_{b}^{-} \to \Xi^{-}\eta_{1})\\
 \mathcal{A}(\Xi_{b}^{0} \to \Xi^{0}\eta_{1})\\
\end{pmatrix}=\begin{pmatrix}
 \frac{1}{2} & 0 & \frac{\sqrt{3}}{2} & 0 \\
 0 & \frac{1}{\sqrt{2}} & 0 & \frac{1}{\sqrt{2}} \\
 \frac{\sqrt{\frac{3}{2}}}{2} & \frac{1}{2} & -\frac{1}{2 \sqrt{2}} & -\frac{1}{2} \\
 \frac{\sqrt{\frac{3}{2}}}{2} & -\frac{1}{2} & -\frac{1}{2 \sqrt{2}} & \frac{1}{2} \\
\end{pmatrix}
\begin{pmatrix}
\langle \mathbf{8} \parallel \mathbf{3}_{0} \parallel \overline{3} \rangle \\
\langle \mathbf{8} \parallel \overline{\mathbf{6}}_{0} \parallel \overline{3} \rangle \\
\langle \mathbf{8} \parallel \mathbf{15}_{0} \parallel \overline{3} \rangle \\
\langle \mathbf{8} \parallel \mathbf{15}_{1} \parallel \overline{3} \rangle
\end{pmatrix}
\end{equation}
The $\Delta S=-1$ decay amplitudes for $\overline{\textbf{3}}_{\mathcal{B}_{b}}\to \textbf{1}_{\mathcal{B}}\, 
\textbf{8}_{\mathcal{M}}$   are $SU(3)$ decomposed in the following way;
\begin{equation}
\begin{pmatrix}
\mathcal{A}(\Lambda_{b}^{0}\to \Lambda_{s}^{0*}\pi^{0})\\
\mathcal{A}(\Lambda_{b}^{0}\to \Lambda_{s}^{0*}\eta_{8})\\
\mathcal{A}(\Xi_{b}^{-}\to \Lambda_{s}^{0*}K^{-})\\
\mathcal{A}(\Xi_{b}^{0}\to \Lambda_{s}^{0*}\overline{K^{0}})\\
\end{pmatrix}
=\begin{pmatrix}
0 & \frac{1}{\sqrt{2}} & 0 & \frac{1}{\sqrt{2}} \\
 \frac{1}{2} & 0 & \frac{\sqrt{3}}{2} & 0 \\
 \frac{\sqrt{\frac{3}{2}}}{2} & \frac{1}{2} & -\frac{1}{2 \sqrt{2}} & -\frac{1}{2} \\
 \frac{\sqrt{\frac{3}{2}}}{2} & -\frac{1}{2} & -\frac{1}{2 \sqrt{2}} & \frac{1}{2} \\
\end{pmatrix}
\begin{pmatrix}
\langle \mathbf{8} \parallel \mathbf{3}_{0} \parallel \overline{3} \rangle \\
\langle \mathbf{8} \parallel \overline{\mathbf{6}}_{0} \parallel \overline{3} \rangle \\
\langle \mathbf{8} \parallel \mathbf{15}_{0} \parallel \overline{3} \rangle \\
\langle \mathbf{8} \parallel \mathbf{15}_{1} \parallel \overline{3} \rangle
\end{pmatrix}
\end{equation}

The $\Delta S=-1$ decay amplitudes for $\overline{\textbf{3}}_{\mathcal{B}_{b}}\to \textbf{8}_{\mathcal{B}}\, 
\textbf{8}_{\mathcal{M}}$ are $SU(3)$ decomposed in the following way;
\begin{equation}
\centering
\rotatebox{90}{
$
\resizebox{\vsize}{!}{%
$
\setcounter{MaxMatrixCols}{22}
\begin{pmatrix}
\mathcal{A}(\Lambda_{b}^{0}\to\Sigma^{0}\eta_{8})\\ \mathcal{A}(\Lambda_{b}^{0}\to \Lambda^{0}\eta_{8})\\ \mathcal{A}(\Lambda_{b}^{0}\to\Sigma^{+}\pi^{-})\\ 
\mathcal{A}(\Lambda_{b}^{0}\to\Sigma^{-}\pi^{+})\\ \mathcal{A}(\Lambda_{b}^{0}\to\Lambda^{0}\pi^{0})\\ \mathcal{A}(\Lambda_{b}^{0}\to\Sigma^{0}\pi^{0})\\ \mathcal{A}(\Lambda_{b}^{0}\to p^{+}K^{-})\\ 
\mathcal{A}(\Lambda_{b}^{0}\to n\bar{K}^{0})\\ \mathcal{A}(\Lambda_{b}^{0}\to\Xi^{-}K^{+})\\ \mathcal{A}(\Lambda_{b}^{0}\to \Xi^{0}K^{0})\\ \mathcal{A}(\Xi_{b}^{-}\to \Xi^{0}\pi^{-})\\
\mathcal{A}(\Xi_{b}^{-}\to \Sigma^{-}\bar{K}^{0})\\ \mathcal{A}(\Xi_{b}^{-}\to\Xi^{-}\eta_{8})\\ \mathcal{A}(\Xi_{b}^{-}\to \Lambda^{0}K^{-})\\ \mathcal{A}(\Xi_{b}^{-}\to \Xi^{-}\pi^{0})\\
\mathcal{A}(\Xi_{b}^{-}\to\Sigma^{0}K^{-})\\ \mathcal{A}(\Xi_{b}^{0}\to\Xi^{-}\pi^{+})\\ \mathcal{A}(\Xi_{b}^{0}\to\Sigma^{0}\bar{K}^{0})\\ \mathcal{A}(\Xi_{b}^{0}\to\Sigma^{+}K^{-})\\
\mathcal{A}(\Xi_{b}^{0}\to\Xi^{0}\eta_{8})\\ \mathcal{A}(\Xi_{b}^{0}\to\Lambda^{0}\bar{K}^{0})\\ \mathcal{A}(\Xi_{b}^{0}\to\Xi^{0}\pi^{0})
\end{pmatrix}
=\begin{pmatrix}
 \frac{1}{2 \sqrt{2}} & \frac{1}{2 \sqrt{2}} & \frac{1}{\sqrt{15}} & \frac{1}{\sqrt{15}} & \frac{1}{\sqrt{6}} & \frac{1}{\sqrt{5}} &
   \frac{1}{2 \sqrt{5}} & \frac{1}{\sqrt{10}} & \frac{1}{\sqrt{10}} & 0 & 0 & 0 & 0 & 0 & 0 & 0 & 0 & 0 & 0 & 0 & 0 & 0 \\
 0 & 0 & 0 & 0 & 0 & 0 & 0 & 0 & 0 & -\frac{1}{2 \sqrt{2}} & \frac{3}{2 \sqrt{10}} & \frac{3}{2 \sqrt{5}} & -\frac{1}{2 \sqrt{5}} &
   -\frac{\sqrt{\frac{3}{5}}}{2} & 0 & 0 & 0 & 0 & 0 & 0 & 0 & 0 \\
 \frac{1}{2 \sqrt{6}} & \frac{1}{2 \sqrt{6}} & 0 & 0 & 0 & -\frac{1}{\sqrt{15}} & -\frac{1}{2 \sqrt{15}} & 0 & 0 & \frac{1}{2 \sqrt{2}}
   & -\frac{1}{6 \sqrt{10}} & -\frac{1}{6 \sqrt{5}} & -\frac{1}{2 \sqrt{5}} & -\frac{\sqrt{\frac{3}{5}}}{2} & \frac{1}{3} & \frac{1}{3
   \sqrt{2}} & \frac{1}{\sqrt{6}} & \frac{1}{\sqrt{6}} & 0 & 0 & 0 & 0 \\
 -\frac{1}{2 \sqrt{6}} & -\frac{1}{2 \sqrt{6}} & 0 & 0 & 0 & \frac{1}{\sqrt{15}} & \frac{1}{2 \sqrt{15}} & 0 & 0 & \frac{1}{2 \sqrt{2}}
   & -\frac{1}{6 \sqrt{10}} & -\frac{1}{6 \sqrt{5}} & -\frac{1}{2 \sqrt{5}} & -\frac{\sqrt{\frac{3}{5}}}{2} & \frac{1}{3} & \frac{1}{3
   \sqrt{2}} & -\frac{1}{\sqrt{6}} & -\frac{1}{\sqrt{6}} & 0 & 0 & 0 & 0 \\
 -\frac{1}{2 \sqrt{2}} & -\frac{1}{2 \sqrt{2}} & \frac{1}{\sqrt{15}} & \frac{1}{\sqrt{15}} & \frac{1}{\sqrt{6}} & -\frac{1}{\sqrt{5}} &
   -\frac{1}{2 \sqrt{5}} & \frac{1}{\sqrt{10}} & \frac{1}{\sqrt{10}} & 0 & 0 & 0 & 0 & 0 & 0 & 0 & 0 & 0 & 0 & 0 & 0 & 0 \\
 0 & 0 & 0 & 0 & 0 & 0 & 0 & 0 & 0 & -\frac{1}{2 \sqrt{2}} & \frac{1}{6 \sqrt{10}} & \frac{1}{6 \sqrt{5}} & \frac{1}{2 \sqrt{5}} &
   \frac{\sqrt{\frac{3}{5}}}{2} & \frac{2}{3} & \frac{\sqrt{2}}{3} & 0 & 0 & 0 & 0 & 0 & 0 \\
 -\frac{1}{2 \sqrt{6}} & -\frac{1}{2 \sqrt{6}} & \frac{1}{3 \sqrt{5}} & \frac{1}{3 \sqrt{5}} & \frac{1}{3 \sqrt{2}} &
   \frac{1}{\sqrt{15}} & \frac{1}{2 \sqrt{15}} & -\frac{\sqrt{\frac{3}{10}}}{2} & -\frac{\sqrt{\frac{3}{10}}}{2} & \frac{1}{2 \sqrt{2}}
   & \frac{1}{2 \sqrt{10}} & \frac{1}{2 \sqrt{5}} & \frac{1}{4 \sqrt{5}} & \frac{\sqrt{\frac{3}{5}}}{4} & 0 & 0 & \frac{1}{2 \sqrt{6}} &
   \frac{1}{2 \sqrt{6}} & \frac{1}{4} & \frac{\sqrt{3}}{4} & 0 & 0 \\
 -\frac{1}{2 \sqrt{6}} & -\frac{1}{2 \sqrt{6}} & \frac{1}{3 \sqrt{5}} & \frac{1}{3 \sqrt{5}} & \frac{1}{3 \sqrt{2}} &
   \frac{1}{\sqrt{15}} & \frac{1}{2 \sqrt{15}} & -\frac{\sqrt{\frac{3}{10}}}{2} & -\frac{\sqrt{\frac{3}{10}}}{2} & -\frac{1}{2 \sqrt{2}}
   & -\frac{1}{2 \sqrt{10}} & -\frac{1}{2 \sqrt{5}} & -\frac{1}{4 \sqrt{5}} & -\frac{\sqrt{\frac{3}{5}}}{4} & 0 & 0 & \frac{1}{2
   \sqrt{6}} & \frac{1}{2 \sqrt{6}} & -\frac{1}{4} & -\frac{\sqrt{3}}{4} & 0 & 0 \\
 \frac{1}{2 \sqrt{6}} & \frac{1}{2 \sqrt{6}} & \frac{1}{3 \sqrt{5}} & \frac{1}{3 \sqrt{5}} & \frac{1}{3 \sqrt{2}} & -\frac{1}{\sqrt{15}}
   & -\frac{1}{2 \sqrt{15}} & -\frac{\sqrt{\frac{3}{10}}}{2} & -\frac{\sqrt{\frac{3}{10}}}{2} & \frac{1}{2 \sqrt{2}} & \frac{1}{2
   \sqrt{10}} & \frac{1}{2 \sqrt{5}} & \frac{1}{4 \sqrt{5}} & \frac{\sqrt{\frac{3}{5}}}{4} & 0 & 0 & -\frac{1}{2 \sqrt{6}} & -\frac{1}{2
   \sqrt{6}} & -\frac{1}{4} & -\frac{\sqrt{3}}{4} & 0 & 0 \\
 \frac{1}{2 \sqrt{6}} & \frac{1}{2 \sqrt{6}} & \frac{1}{3 \sqrt{5}} & \frac{1}{3 \sqrt{5}} & \frac{1}{3 \sqrt{2}} & -\frac{1}{\sqrt{15}}
   & -\frac{1}{2 \sqrt{15}} & -\frac{\sqrt{\frac{3}{10}}}{2} & -\frac{\sqrt{\frac{3}{10}}}{2} & -\frac{1}{2 \sqrt{2}} & -\frac{1}{2
   \sqrt{10}} & -\frac{1}{2 \sqrt{5}} & -\frac{1}{4 \sqrt{5}} & -\frac{\sqrt{\frac{3}{5}}}{4} & 0 & 0 & -\frac{1}{2 \sqrt{6}} &
   -\frac{1}{2 \sqrt{6}} & \frac{1}{4} & \frac{\sqrt{3}}{4} & 0 & 0 \\
 \frac{1}{2 \sqrt{6}} & -\frac{1}{2 \sqrt{6}} & \frac{1}{2 \sqrt{5}} & -\frac{1}{2 \sqrt{5}} & 0 & \frac{1}{2 \sqrt{15}} &
   -\frac{1}{\sqrt{15}} & \frac{\sqrt{\frac{3}{10}}}{2} & -\frac{\sqrt{\frac{3}{10}}}{2} & 0 & -\frac{1}{3 \sqrt{10}} & \frac{1}{6
   \sqrt{5}} & -\frac{3}{4 \sqrt{5}} & \frac{\sqrt{\frac{3}{5}}}{4} & \frac{1}{6} & -\frac{1}{3 \sqrt{2}} & -\frac{1}{2 \sqrt{6}} &
   \frac{1}{2 \sqrt{6}} & \frac{1}{4} & -\frac{1}{4 \sqrt{3}} & -\frac{1}{2 \sqrt{3}} & \frac{1}{2 \sqrt{3}} \\
 -\frac{1}{2 \sqrt{6}} & \frac{1}{2 \sqrt{6}} & \frac{1}{2 \sqrt{5}} & -\frac{1}{2 \sqrt{5}} & 0 & -\frac{1}{2 \sqrt{15}} &
   \frac{1}{\sqrt{15}} & \frac{\sqrt{\frac{3}{10}}}{2} & -\frac{\sqrt{\frac{3}{10}}}{2} & 0 & -\frac{1}{3 \sqrt{10}} & \frac{1}{6
   \sqrt{5}} & -\frac{3}{4 \sqrt{5}} & \frac{\sqrt{\frac{3}{5}}}{4} & \frac{1}{6} & -\frac{1}{3 \sqrt{2}} & \frac{1}{2 \sqrt{6}} &
   -\frac{1}{2 \sqrt{6}} & -\frac{1}{4} & \frac{1}{4 \sqrt{3}} & \frac{1}{2 \sqrt{3}} & -\frac{1}{2 \sqrt{3}} \\
 \frac{1}{4} & -\frac{1}{4} & \frac{1}{2 \sqrt{30}} & \sqrt{\frac{2}{15}} & -\frac{1}{2 \sqrt{3}} & 0 & 0 & \frac{1}{4 \sqrt{5}} &
   -\frac{1}{4 \sqrt{5}} & 0 & \frac{\sqrt{\frac{3}{5}}}{2} & -\frac{\sqrt{\frac{3}{10}}}{2} & -\frac{\sqrt{\frac{3}{10}}}{4} &
   \frac{1}{4 \sqrt{10}} & 0 & 0 & \frac{1}{4} & -\frac{1}{4} & -\frac{\sqrt{\frac{3}{2}}}{4} & \frac{1}{4 \sqrt{2}} & -\frac{1}{2
   \sqrt{2}} & 0 \\
 -\frac{1}{4} & \frac{1}{4} & \frac{1}{2 \sqrt{30}} & \sqrt{\frac{2}{15}} & -\frac{1}{2 \sqrt{3}} & 0 & 0 & \frac{1}{4 \sqrt{5}} &
   -\frac{1}{4 \sqrt{5}} & 0 & \frac{\sqrt{\frac{3}{5}}}{2} & -\frac{\sqrt{\frac{3}{10}}}{2} & -\frac{\sqrt{\frac{3}{10}}}{4} &
   \frac{1}{4 \sqrt{10}} & 0 & 0 & -\frac{1}{4} & \frac{1}{4} & \frac{\sqrt{\frac{3}{2}}}{4} & -\frac{1}{4 \sqrt{2}} & \frac{1}{2
   \sqrt{2}} & 0 \\
 -\frac{1}{4 \sqrt{3}} & \frac{1}{4 \sqrt{3}} & \frac{7}{6 \sqrt{10}} & -\frac{1}{3 \sqrt{10}} & -\frac{1}{6} & \frac{1}{\sqrt{30}} &
   -\sqrt{\frac{2}{15}} & -\frac{\sqrt{\frac{3}{5}}}{4} & \frac{\sqrt{\frac{3}{5}}}{4} & 0 & \frac{1}{6 \sqrt{5}} & -\frac{1}{6
   \sqrt{10}} & \frac{3}{4 \sqrt{10}} & -\frac{\sqrt{\frac{3}{10}}}{4} & \frac{1}{3 \sqrt{2}} & -\frac{1}{3} & \frac{1}{4 \sqrt{3}} &
   -\frac{1}{4 \sqrt{3}} & -\frac{1}{4 \sqrt{2}} & \frac{1}{4 \sqrt{6}} & \frac{1}{2 \sqrt{6}} & \frac{1}{\sqrt{6}} \\
 \frac{1}{4 \sqrt{3}} & -\frac{1}{4 \sqrt{3}} & \frac{7}{6 \sqrt{10}} & -\frac{1}{3 \sqrt{10}} & -\frac{1}{6} & -\frac{1}{\sqrt{30}} &
   \sqrt{\frac{2}{15}} & -\frac{\sqrt{\frac{3}{5}}}{4} & \frac{\sqrt{\frac{3}{5}}}{4} & 0 & \frac{1}{6 \sqrt{5}} & -\frac{1}{6
   \sqrt{10}} & \frac{3}{4 \sqrt{10}} & -\frac{\sqrt{\frac{3}{10}}}{4} & \frac{1}{3 \sqrt{2}} & -\frac{1}{3} & -\frac{1}{4 \sqrt{3}} &
   \frac{1}{4 \sqrt{3}} & \frac{1}{4 \sqrt{2}} & -\frac{1}{4 \sqrt{6}} & -\frac{1}{2 \sqrt{6}} & -\frac{1}{\sqrt{6}} \\
 \frac{1}{2 \sqrt{6}} & -\frac{1}{2 \sqrt{6}} & \frac{1}{2 \sqrt{5}} & -\frac{1}{2 \sqrt{5}} & 0 & \frac{1}{2 \sqrt{15}} &
   -\frac{1}{\sqrt{15}} & \frac{\sqrt{\frac{3}{10}}}{2} & -\frac{\sqrt{\frac{3}{10}}}{2} & 0 & \frac{1}{3 \sqrt{10}} & -\frac{1}{6
   \sqrt{5}} & \frac{3}{4 \sqrt{5}} & -\frac{\sqrt{\frac{3}{5}}}{4} & -\frac{1}{6} & \frac{1}{3 \sqrt{2}} & -\frac{1}{2 \sqrt{6}} &
   \frac{1}{2 \sqrt{6}} & -\frac{1}{4} & \frac{1}{4 \sqrt{3}} & \frac{1}{2 \sqrt{3}} & -\frac{1}{2 \sqrt{3}} \\
 \frac{1}{4 \sqrt{3}} & -\frac{1}{4 \sqrt{3}} & \frac{7}{6 \sqrt{10}} & -\frac{1}{3 \sqrt{10}} & -\frac{1}{6} & -\frac{1}{\sqrt{30}} &
   \sqrt{\frac{2}{15}} & -\frac{\sqrt{\frac{3}{5}}}{4} & \frac{\sqrt{\frac{3}{5}}}{4} & 0 & -\frac{1}{6 \sqrt{5}} & \frac{1}{6
   \sqrt{10}} & -\frac{3}{4 \sqrt{10}} & \frac{\sqrt{\frac{3}{10}}}{4} & -\frac{1}{3 \sqrt{2}} & \frac{1}{3} & -\frac{1}{4 \sqrt{3}} &
   \frac{1}{4 \sqrt{3}} & -\frac{1}{4 \sqrt{2}} & \frac{1}{4 \sqrt{6}} & \frac{1}{2 \sqrt{6}} & \frac{1}{\sqrt{6}} \\
 -\frac{1}{2 \sqrt{6}} & \frac{1}{2 \sqrt{6}} & \frac{1}{2 \sqrt{5}} & -\frac{1}{2 \sqrt{5}} & 0 & -\frac{1}{2 \sqrt{15}} &
   \frac{1}{\sqrt{15}} & \frac{\sqrt{\frac{3}{10}}}{2} & -\frac{\sqrt{\frac{3}{10}}}{2} & 0 & \frac{1}{3 \sqrt{10}} & -\frac{1}{6
   \sqrt{5}} & \frac{3}{4 \sqrt{5}} & -\frac{\sqrt{\frac{3}{5}}}{4} & -\frac{1}{6} & \frac{1}{3 \sqrt{2}} & \frac{1}{2 \sqrt{6}} &
   -\frac{1}{2 \sqrt{6}} & \frac{1}{4} & -\frac{1}{4 \sqrt{3}} & -\frac{1}{2 \sqrt{3}} & \frac{1}{2 \sqrt{3}} \\
 -\frac{1}{4} & \frac{1}{4} & -\frac{1}{2 \sqrt{30}} & -\sqrt{\frac{2}{15}} & \frac{1}{2 \sqrt{3}} & 0 & 0 & -\frac{1}{4 \sqrt{5}} &
   \frac{1}{4 \sqrt{5}} & 0 & \frac{\sqrt{\frac{3}{5}}}{2} & -\frac{\sqrt{\frac{3}{10}}}{2} & -\frac{\sqrt{\frac{3}{10}}}{4} &
   \frac{1}{4 \sqrt{10}} & 0 & 0 & -\frac{1}{4} & \frac{1}{4} & -\frac{\sqrt{\frac{3}{2}}}{4} & \frac{1}{4 \sqrt{2}} & -\frac{1}{2
   \sqrt{2}} & 0 \\
 \frac{1}{4} & -\frac{1}{4} & -\frac{1}{2 \sqrt{30}} & -\sqrt{\frac{2}{15}} & \frac{1}{2 \sqrt{3}} & 0 & 0 & -\frac{1}{4 \sqrt{5}} &
   \frac{1}{4 \sqrt{5}} & 0 & \frac{\sqrt{\frac{3}{5}}}{2} & -\frac{\sqrt{\frac{3}{10}}}{2} & -\frac{\sqrt{\frac{3}{10}}}{4} &
   \frac{1}{4 \sqrt{10}} & 0 & 0 & \frac{1}{4} & -\frac{1}{4} & \frac{\sqrt{\frac{3}{2}}}{4} & -\frac{1}{4 \sqrt{2}} & \frac{1}{2
   \sqrt{2}} & 0 \\
 -\frac{1}{4 \sqrt{3}} & \frac{1}{4 \sqrt{3}} & \frac{7}{6 \sqrt{10}} & -\frac{1}{3 \sqrt{10}} & -\frac{1}{6} & \frac{1}{\sqrt{30}} &
   -\sqrt{\frac{2}{15}} & -\frac{\sqrt{\frac{3}{5}}}{4} & \frac{\sqrt{\frac{3}{5}}}{4} & 0 & -\frac{1}{6 \sqrt{5}} & \frac{1}{6
   \sqrt{10}} & -\frac{3}{4 \sqrt{10}} & \frac{\sqrt{\frac{3}{10}}}{4} & -\frac{1}{3 \sqrt{2}} & \frac{1}{3} & \frac{1}{4 \sqrt{3}} &
   -\frac{1}{4 \sqrt{3}} & \frac{1}{4 \sqrt{2}} & -\frac{1}{4 \sqrt{6}} & -\frac{1}{2 \sqrt{6}} & -\frac{1}{\sqrt{6}} \\
\end{pmatrix}
\begin{pmatrix}
\langle 10 \parallel \mathbf{15_{1}}\parallel \bar{3}\rangle\\ 
\langle 10 \parallel \mathbf{15^{'}_{1}} \parallel \bar{3} \rangle \\ 
\langle 27\parallel \mathbf{15_{1}} \parallel \bar{3}\rangle \\
\langle 27\parallel \mathbf{24_{1}}\parallel \bar{3}\rangle\\ 
\langle 27\parallel\mathbf{42_{1}}\parallel \bar{3}\rangle\\ 
\langle \bar{10}\parallel \mathbf{24_{1}} \parallel \bar{3}\rangle\\ 
\langle\bar{10} \parallel \mathbf{\bar{6}_{1}} \parallel \bar{3}\rangle\\
\langle 8_{1}\parallel \mathbf{15_{1}} \parallel \bar{3}\rangle\\ 
\langle 8_{1}\parallel \mathbf{\bar{6}_{1}} \parallel \bar{3}\rangle\\ 
\langle 1\parallel \mathbf{3_{0}}\parallel \bar{3}\rangle\\ 
\langle 27 \parallel \mathbf{15_{0}}\parallel \bar{3}\rangle\\ 
\langle 27\parallel \mathbf{42_{0}}\parallel \bar{3}\rangle\\ 
\langle 8_{1} \parallel \mathbf{3_ {0}} \parallel \bar{3}\rangle\\ 
\langle 8_{1}\parallel\mathbf{15_{0}}\parallel \bar{3}\rangle\\ 
\langle 27\parallel \mathbf{24_{2}} \parallel \bar{3}\rangle\\ 
\langle 27\parallel \mathbf{42_{2}}\parallel \bar{3}\rangle\\ 
\langle 8_{2}\parallel \mathbf{15_{1}} \parallel \bar{3}\rangle\\ 
\langle 8_{2}\parallel \mathbf{\bar{6}_{1}} \parallel \bar{3}\rangle\\ 
\langle 8_{2} \parallel \mathbf{3_{0}}\parallel \bar{3}\rangle\\ 
\langle 8_{2}\parallel\mathbf{15_{0}}\parallel \bar{3}\rangle\\ 
\langle 10 \parallel \mathbf{15_{0}} \parallel \bar{3}\rangle\\ 
\langle \bar{10}\parallel \mathbf{24_{2}}\parallel \bar{3}\rangle 
\end{pmatrix}
$
	}
$
	}
\nonumber	
\end{equation}
\end{widetext}%\onecolumngrid
\onecolumngrid

%%%%%%%%%%%%%%%%%%%%%%%%%%%%%%%
\twocolumngrid

\end{document}